\documentclass[a4paper,11pt]{article}
\pdfoutput=1
\usepackage{jcappub}
\usepackage[T1]{fontenc} % if needed
\usepackage{hyperref}
\usepackage{graphics}
\usepackage{amssymb, amsmath}
\usepackage{bm}
\usepackage{xcolor,color,colortbl}
\usepackage{multirow,bigdelim}
\usepackage{bbold}
\usepackage{verbatim}
\usepackage{booktabs}
\usepackage{graphicx}
\usepackage{subcaption}
\usepackage{caption}
\usepackage[nameinlink,capitalise]{cleveref}
\usepackage{lineno}
\newcommand{\fnl}{\(f_{\mathrm{NL}}\)}
\newcommand{\Planck}{\textit{Planck }}  % or aastex63
 
\usepackage{orcidlink}

\author[1,2,3]{S.~Chiarenza\, \orcidlink{0009-0003-6369-9904}}
\author[1,2,3]{A.~Krolewski}
\author[1,2,3]{M.~Bonici}
\author[4]{E.~Chaussidon\,\orcidlink{0000-0001-8996-4874}}
\author[4]{R.~de Belsunce\,\orcidlink{0000-0003-3660-4028}}
\author[1,2,3]{W.~J.~Percival\,\orcidlink{0000-0002-0644-5727}}
\author[4]{J.~Aguilar}
\author[6]{S.~Ahlen\,\orcidlink{0000-0001-6098-7247}}
\author[4,5]{A.~Baleato Lizancos\,\orcidlink{0000-0002-0232-6480}}
\author[7,8]{D.~Bianchi\,\orcidlink{0000-0001-9712-0006}}
\author[9]{D.~Brooks}
\author[4]{T.~Claybaugh}
\author[4]{A.~Cuceu\,\orcidlink{0000-0002-2169-0595}}
\author[10]{K.~S.~Dawson\,\orcidlink{0000-0002-0553-3805}}
\author[11]{A.~de la Macorra\,\orcidlink{0000-0002-1769-1640}}
\author[9]{P.~Doel}
\author[4,5]{S.~Ferraro\,\orcidlink{0000-0003-4992-7854}}
\author[13]{A.~Font-Ribera\,\orcidlink{0000-0002-3033-7312}}
\author[14,15]{J.~E.~Forero-Romero\,\orcidlink{0000-0002-2890-3725}}
\author[16,17,18]{E.~Gaztañaga\,\orcidlink{0000-0001-9632-0815}}
\author[4,19]{S.~Gontcho A Gontcho\,\orcidlink{0000-0003-3142-233X}}
\author[20]{G.~Gutierrez}
\author[21,23]{H.~K.~Herrera-Alcantar\,\orcidlink{0000-0002-9136-9609}}
\author[23,24,25]{K.~Honscheid\,\orcidlink{0000-0002-6550-2023}}
\author[26,27]{D.~Huterer\,\orcidlink{0000-0001-6558-0112}}
\author[28]{M.~Ishak\,\orcidlink{0000-0002-6024-466X}}
\author[29]{R.~Joyce\,\orcidlink{0000-0003-0201-5241}}
\author[30]{D.~Kirkby\,\orcidlink{0000-0002-8828-5463}}
\author[4]{A.~Kremin\,\orcidlink{0000-0001-6356-7424}}
\author[9]{O.~Lahav}
\author[25]{C.~Lamman\,\orcidlink{0000-0002-6731-9329}}
\author[4]{M.~Landriau\,\orcidlink{0000-0003-1838-8528}}
\author[31]{L.~Le~Guillou\,\orcidlink{0000-0001-7178-8868}}
\author[4]{M.~E.~Levi\,\orcidlink{0000-0003-1887-1018}}
\author[32,13]{M.~Manera\,\orcidlink{0000-0003-4962-8934}}
\author[23,25,33]{P.~Martini\,\orcidlink{0000-0002-4279-4182}}
\author[29]{A.~Meisner\,\orcidlink{0000-0002-1125-7384}}
\author[13,34]{R.~Miquel}
\author[17]{S.~Nadathur\,\orcidlink{0000-0001-9070-3102}}
\author[35]{J.~ A.~Newman\,\orcidlink{0000-0001-8684-2222}}
\author[36,37]{G.~Niz\,\orcidlink{0000-0002-1544-8946}}
\author[22,4]{N.~Palanque-Delabrouille\,\orcidlink{0000-0003-3188-784X}}
\author[4,38,5]{C.~Poppett}
\author[39]{F.~Prada\,\orcidlink{0000-0001-7145-8674}}
\author[40]{I.~P\'erez-R\`afols\,\orcidlink{0000-0001-6979-0125}}
\author[41]{G.~Rossi}
\author[42]{E.~Sanchez\,\orcidlink{0000-0002-9646-8198}}
\author[4]{D.~Schlegel}
\author[26,27]{M.~Schubnell}
\author[43]{H.~Seo\,\orcidlink{0000-0002-6588-3508}}
\author[4]{J.~Silber\,\orcidlink{0000-0002-3461-0320}}
\author[29]{D.~Sprayberry}
\author[27]{G.~Tarl\'{e}\,\orcidlink{0000-0003-1704-0781}}
\author[29]{B.~A.~Weaver}
\author[22]{C.~Yèche\,\orcidlink{0000-0001-5146-8533}}
\author[4]{R.~Zhou\,\orcidlink{0000-0001-5381-4372}}
\author[44]{H.~Zou\,\orcidlink{0000-0002-6684-3997}}

\affiliation[1]{Department of Physics and Astronomy, University of Waterloo, 200 University Ave W, Waterloo, ON N2L 3G1, Canada}
\affiliation[2]{Perimeter Institute for Theoretical Physics, 31 Caroline St. North, Waterloo, ON N2L 2Y5, Canada}
\affiliation[3]{Waterloo Centre for Astrophysics, University of Waterloo, 200 University Ave W, Waterloo, ON N2L 3G1, Canada}

\affiliation{For other affiliations, see Appendix~\ref{app:affiliations}.}

\emailAdd{schiaren@uwaterloo.ca}

\title{Constraining primordial non-Gaussianity from DESI DR1 quasars and \Planck PR4 CMB Lensing}


\abstract{We present the first measurement of local-type primordial non-Gaussianity from the cross-correlation between $1.2$ million spectroscopically confirmed quasars from the first data release (DR1) of the Dark Energy Spectroscopic Instrument (DESI) and the \Planck PR4 CMB lensing reconstructions. The analysis is performed in three tomographic redshift bins covering $0.8 < z < 3.5$, covering a sky fraction of $\sim 20\%$. We adopt a catalog-based pseudo-$C_\ell$ estimator and apply linear imaging weights validated on noiseless mocks. Compared to previous analyses using photometric quasar samples, our results benefit from the high purity of the DESI spectroscopic sample, the reduced noise of PR4 lensing, and the absence of excess large-scale power in the spectroscopic quasar auto-correlation. Fitting simultaneously for the non-Gaussianity parameter $f_{\mathrm{NL}}$ and the linear bias amplitude in each redshift bin, we obtain $f_{\mathrm{NL}} = 2^{+28}_{-34}$ for a response parameter $p=1.6$, and $f_{\mathrm{NL}} = 6^{+20}_{-24}$ for $p=1.0$.
These results improve the constraints on $f_{\mathrm{NL}}$ by $\sim 35\%$ compared to the previous analysis based on the Legacy Imaging Survey DR9.
Additionally, we derive an optimal weighting scheme to maximize the constraining power. In this case, and assuming $p=1.6$, we obtain $f_\mathrm{NL}=19^{+25}_{-31}$. 
Our results demonstrate the statistical power of DESI quasars for probing inflationary physics, and highlight the promise of future DESI data releases.}
\begin{document}
\maketitle
\flushbottom

\section{Introduction}\label{sec:intro}
Inflation is the leading framework for explaining the emergence of cosmic structure. A key prediction of this theory is that the initial curvature perturbations are almost perfectly Gaussian \citep{bardeen1983spontaneous, falk1992dependence, acquaviva2002second}. This prediction is supported by precise measurements of the Cosmic Microwave Background (CMB) temperature and polarization \citep{akrami2020planck}. Still, many inflation models allow small departures from Gaussianity, constraining these deviations, known as primordial non-Gaussianity (PNG), is a powerful direct test of early-Universe physics. For example, in single-field slow-roll (SFSR) inflation the expected level of local PNG is small, $f_{\mathrm{NL}}^{\mathrm{loc}}\!\sim\!\mathcal{O}(10^{-2})$ \citep{Maldacena03,Acquaviva03}. Multi-field scenarios, on the other hand, can naturally reach $f_{\mathrm{NL}}^{\mathrm{loc}}\!\sim\!\mathcal{O}(1)$ \citep{Bartolo04}.  
Achieving order-unity precision on $f_{\mathrm{NL}}^{\mathrm{loc}}$ is thus an important goal in ruling out significant parts of the parameter space of inflation. 

The best current limit comes from the \Planck bispectrum, $f_{\mathrm{NL}}\!=\!-0.9\pm5.1$ \citep{PlanckFnl}. This is already close to the cosmic-variance limit for the CMB, and future CMB experiments will be able to improve it by only a factor of two \citep{CMBS4-book}. Stronger constraints must therefore come from large-scale structure (LSS). 

Local PNG introduces a scale-dependent bias in the clustering of galaxies \citep{Dalal08}, which manifests as an enhancement of the quasars (QSOs) power spectrum, $P(k)$, on the very large scales. The effect scales as $\propto k^{-2}$ and grows with redshift, making quasars particularly powerful tracers: their high-redshift distribution provides access to the large cosmic volumes and low-$k$ modes where the signal is most pronounced. In fact, the best constraint to date from LSS comes from the 3D power spectrum analysis done on the combination of DESI DR1 QSO and LRG samples, yielding $f_{\mathrm{NL}} = -3.6^{+9.0}_{-9.1}$ \citep{chaussidon2024}.
However, measuring galaxy clustering on very large scales is inherently challenging. The signal fluctuations are small in amplitude, and systematic effects can introduce significant spurious power, particularly on the largest scales \citep{Slosar08,Xia11,Nikoloudakis12,Pullen13,Leistedt_Peiris,Giannantonio14a,Leistedt14,Ho15}. As a result, all large-scale structure constraints on \fnl{} to date have been limited by systematic uncertainties rather than fundamental statistical noise \citep{rezaie2024local}.

Cross-correlations between galaxy surveys and CMB lensing provide a robust alternative to galaxy auto-correlations for probing primordial non-Gaussianity. The main advantage comes from the fact that the dominant sources of noise and systematics in galaxy surveys are generally uncorrelated with those in CMB experiments \citep{Smith2007,Hirata2008,Chang10,Rhodes13}. This property makes cross-correlations especially powerful for controlling large-scale systematics and isolating the true cosmological signal \citep{Rhodes13,Giannantonio14b}. In fact, while systematics can bias the galaxy auto-correlation signal, in the cross-correlation they primarily act to increase the statistical variance without introducing a systematic shift.  We note that, in principle, correlated systematics could exist: galactic emission can bias lensing reconstruction and is correlated with extinction, but no significant effects have been detected so far.
Moreover, the CMB lensing reconstruction can suffer from a potential contamination from extragalactic foregrounds, the main one being the Cosmic Infrared Background (CIB) and thermal Sunyaev-Zel'dovich (tSZ) effect, particularly when the reconstruction is derived from temperature data \citep{lizancos2025halo}. However, such effects are expected to be small for our measurement, and no significant bias has been reported in similar cross-correlation analyses \citep{piccirilli2024growth}. As a supporting test, we performed the measurement using the Planck polarization-only lensing reconstruction (available within the PR4 data release), which is strongly unaffected by CIB and tSZ contamination. While the polarization-only data yield much larger statistical uncertainties, the recovered signal shows no significant deviation from our baseline result.
Another benefit of CMB lensing cross-correlations is their sensitivity to high redshifts, where the \fnl{} signature is strongest. The lensing kernel for CMB lensing peaks at $z\approx 2$, overlapping well with the redshift distribution of quasars and other high-redshift tracers. There have been several recent applications of this idea \citep{bermejo2025constraints, fabbian2025, krolewski2024constraining}.
In particular the constraint from the Quaia photometric quasar sample cross-correlated with \Planck{} PR4 lensing \citep{carron2022cmb} yielded $\sigma_{f_{\mathrm{NL}}} \approx 25$ \citep{fabbian2025}. The reported constraint using DESI LRGs is of $\sigma_{f_{\mathrm{NL}}} \approx 40$ \citep{bermejo2025constraints}. Finally, the analysis of the photometric DESI Legacy Survey quasar sample cross-correlated with \Planck{} PR4 lensing presented in \citep{krolewski2024constraining}, which forms the basis of the approach used in this work, obtained $\sigma_{f_{\mathrm{NL}}} \approx 45$. Looking ahead, forecasts suggest that ongoing spectroscopic surveys like DESI, and future Stage 4 surveys like SPHEREx and LSST, combined with CMB lensing, are advancing toward the era of $\sigma_{f_{\textrm{NL}}} < 1$ \citep{Seljak09,Dore14,Yamauchi14,Karagiannis18,SchmittfullSeljak,Ferraro20,Gualdi21,Schlegel22}.

Driven by this motivation, we present the first measurement of local-type non-Gaussianity from the cross-correlation between DESI DR1 quasars and \Planck PR4 CMB lensing maps \citep{akrami2020planck}. Our analysis focuses exclusively on the cross-correlation to minimize contamination from auto-correlation systematics, providing a clean and complementary test of the standard LSS analysis which employs the 3D power spectrum of high redshift tracers. The analysis builds upon that by \citep{krolewski2024constraining}: DESI DR1 spectroscopically confirmed quasars are employed instead of the photometric targets in the DESI Legacy Survey \citep{Dey19}, so the sample is affected by fewer imaging systematics. \Planck PR4 lensing maps are employed, which have $10-20\%$ less noise than the \Planck 2018 maps \citep{aghanim2020planck} due to use of additional data and more optimal filtering and analysis methods \citep{akrami2020planck, carron2022cmb}. Our constraining power now reaches $\sigma_{f_{\textrm{NL}}} \sim 20$, showing that the purer quasar sample and the new pixel-free pipeline improve the constraining power by a factor $\gtrsim 2$ over \citep{krolewski2024constraining}, who find $\sigma_{f_{\textrm{NL}}} \sim 45$ despite the $\sim 2\times$ larger sky fraction.
 
In Section~\ref{sec:theory}, we describe the theory necessary to compute angular correlation functions at low $\ell$. 
In Section~\ref{sec:data}, we describe the quasar and CMB lensing data, and in Section~\ref{sec:ps}, we describe our angular power spectrum pipeline. In Section~\ref{sec:sys_weights}, we validate our pipeline on mocks, to verify that we do not over-correct when mitigating the effect of imaging systematics. In Sec.~\ref{sec:optw} the implementation of optimal redshift weights to maximize the \fnl{} signal in the cross-correlation is discussed. Finally, in Section~\ref{sec:results}, we present the results, and in Section~\ref{sec:conc} we compare them to previous $f_{\textrm{NL}}$ constraints.
Throughout this paper, we fix the cosmological parameters
to the \Planck 2018  flat $\Lambda$CDM model \citep{PlanckLegacy18} with $h = 0.6766$, $A_\mathrm{s} = 2.105 \times 10^{-9}$, $n_\mathrm{s} = 0.9665$, $\Omega_m = 0.3096$, $\Omega_b = 0.049$, one neutrino with mass $0.06$ eV, and $\sigma_8 = 0.8102$, but we also test the impact of freeing the primordial power spectrum (\textit{i.e.}, $A_\mathrm{s}$ and $n_\mathrm{s}$) on the constraints we obtain on \fnl{}.

\section{Theory}
\label{sec:theory}
Local PNG is characterized by the dimensionless parameter \fnl{}:
\begin{equation}
    \Phi_{\mathrm{NG}}(\mathbf{x}) = \phi_\mathrm{G}(\mathbf{x}) + f_{\mathrm{NL}} \left( \phi_\mathrm{G}^2(\mathbf{x}) - \langle \phi_\mathrm{G}^2 \rangle \right),
\label{eq:fnl_def}
\end{equation}
where \(\phi_\mathrm{G}\) is a Gaussian random field representing the primordial gravitational potential \citep{KomatsuSpergel01}. This local transformation introduces mode-coupling between long and short wavelengths, which manifests as a scale-dependent bias in large-scale structure tracers \citep{Dalal08,Slosar08,SeljakDesjacques}.
The physical origin of PNG-induced scale-dependent bias lies in the modulation of halo formation by long-wavelength potential fluctuations. Taking the Laplacian of Eq.~\ref{eq:fnl_def}, we see that the presence of PNG increases the density in the peaks of the density field: 
\begin{equation}
    \delta_{\mathrm{NG}} = \delta_{\mathrm{G}} (1 + 2f_{\mathrm{NL}}\phi_\mathrm{G})
\end{equation}
This shifts the effective collapse threshold: 
\begin{equation}
    \delta_c \rightarrow \delta_c (1- 2f_{\mathrm{NL}}\phi_\mathrm{G})
\end{equation}
And, following \citep{Dalal08}, one can show that the resulting scale-dependent correction to the linear bias \(b_1(z)\) is:
\begin{equation}\label{eqn:bias_total}
    b(k,z) = b_1(z) + \frac{b_\Phi(z)}{T_{\Phi\rightarrow\delta}(k,z)}f_{\mathrm{NL}},
\end{equation}
where $b_1(z)$ is the linear bias of the tracer, $b_\Phi$ is the PNG bias, giving the response to the presence of local PNG of the tracer, and the transfer function \( T_{\Phi\rightarrow\delta}(k,z)\) is the transfer function between the primordial gravitational field $\phi_\mathrm{G}$ and the matter density perturbation, computed as:
\begin{equation}\label{eqn:pk_def}
    T_{\Phi \rightarrow \delta}(k, z)=\sqrt{\frac{P_{\text {lin }}(k, z)}{P_{\Phi}(k)}} \quad \text { with } \quad P_{\Phi}(k)=\frac{9}{25} \frac{2 \pi^2}{k^3} A_\mathrm{s}\left(\frac{k}{k_{\text {pivot }}}\right)^{n_\mathrm{s}-1},
\end{equation}
Hence, through the Poisson equation, $T_{\Phi \rightarrow \delta}(k, z)$ has the well-known scale dependence:
\begin{equation}
    T_{\Phi \rightarrow \delta}(k, z) \propto \, k^2 \, T_{\Phi \rightarrow \Phi}(k, z)
\end{equation}
where $ T_{\Phi \rightarrow \Phi}(k, z)$ is the usual total matter transfer function.
Since \fnl{} is always paired with $b_\Phi$, a measurement of scale-dependent bias measures the product $b_\Phi f_{\mathrm{NL}}$. Nonetheless, this degeneracy does not affect the significance of a potential non-zero detection, which would remain a robust indication of the presence of PNG \citep{barreira2022can, barreira2020impact}.
The PNG bias \( b_\Phi \) quantifies the logarithmic response of the galaxy number density to a change in amplitude of the matter clustering. It is defined as:
\begin{equation}\label{eqn:bphi_theory}
    b_\Phi=\frac{\partial \log \,\bar{n}}{\partial \log \, \sigma_8}.
\end{equation}
However, the theoretical modeling of $b_\Phi$ is a widely discussed topic \citep{biagetti2017verifying, barreira2020impact, barreira2022can, sullivan2023learning} that goes beyond the scope of this present work. Therefore, we will follow the standard prescription, also known as universality relation \citep{slosar2008constraints}:
\begin{equation}\label{eqn:bphi_sims}
    b_\Phi(z) = 2\delta_c(b_1(z)-p)
\end{equation}
where \(\delta_c \approx 1.686\) is the critical spherical collapse density threshold and \(b_1(z)\) is the redshift-dependent linear bias. The most typical choices for the response parameter $p$ is $1$ for a mass-selected sample, while for a sample dominated by recent mergers, such as quasars, $p = 1.6$ is a more appropriate choice \citep{slosar2008constraints, breiding2024powerful}. We will report constraints for both prescriptions. 

In this work, we probe $\Delta b(k)$ using the matter-galaxy cross-power spectrum, $P_{gm}(k)$. Specifically, since CMB lensing is a 2-dimensional projected field, our observable is the angular cross-power spectrum:
\begin{equation}
C_{\ell}^{\kappa \delta} = \frac{2}{\pi} \int \mathrm{d}z_1 W_\delta(z_1) \int \mathrm{d}z_2 \frac{W_\kappa(z_2)}{\chi^2(z_2)}
\int \mathrm{d}k \,P_{mm}(k, z_1, z_2) j_\ell(k\chi(z_1)) j_\ell(k\chi(z_2))
\label{eqn:gg_full}
\end{equation}
where $W_\delta(z)$ is the number counts kernel:
\begin{equation}\label{eqn:number_counts}
    W_\delta(z) = \frac{H(z)}{c}b_1(z)n(z),
\end{equation}
$n(z)$ being the quasar target redshift distribution, and $W_\kappa(z)$ is the CMB lensing kernel
\begin{equation}\label{eqn:cmb_kernel}
   W_\kappa(z) =  \frac{3}{2}\frac{H_0^2}{c^2}\Omega_m \,\chi(z)(1+z)\left(1-\frac{\chi(z)}{\chi(z_\star)}\right)
\end{equation}
with $\chi(z_\star)$ the comoving distance to the surface of last scattering, at $z_\star = 1090$.
When including the scale-dependent bias induced by $f_{\textrm{NL}}$, we replace the linear galaxy bias $b(z)$ with a scale- and redshift-dependent term $b(k, z)$, as defined in Eq.~\ref{eqn:bias_total}. This assumes a linear galaxy bias, valid over the scales used here ($4 < \ell < 300$, i.e.\ $ k \lesssim 0.1~h\,\mathrm{Mpc}^{-1}$ for $0.8 < z < 3.5$).
 
In accordance with other PNG analysis within DESI, we fix the redshift evolution of the bias to follow the functional form of \citep{chaussidon2024},
%\begin{equation}\label{eqn:chaussidon_bias}
%    b(z) = b_0^i \left[0.237(1 + z)^2 + 0.771\right],
%\end{equation}
which captures the expected increase of quasar bias with redshift. In practice, this redshift dependence is integrated over in the model, and therefore modifies the effective redshift at which the measurement is interpreted. We introduce a free normalization parameter $b_0^i$ for each tomographic bin, which preserves the redshift evolution but allows the overall amplitude to vary, resulting in discontinuities between bins. In Sec.~\ref{sec:bias_test} we demonstrate that \fnl{} depends only on the effective bias amplitude of each bin, as the constraints remain consistent across all alternative bias evolution models considered.

Eq.~\ref{eqn:gg_full} is a numerically challenging three-dimensional integral over oscillatory spherical Bessel functions. The Limber approximation, typically employed to simplify the problem, approximates the spherical Bessel functions as Dirac delta functions in their first peak \citep{Limber53}:
\begin{equation}
    j_\ell(k \chi) \rightarrow
\sqrt{\frac{\pi}{2 \ell + 1}} \delta_D(\ell + 1/2 - k\chi).
\end{equation}
The three-dimensional integral is reduced to a one-dimensional integral in $k$ or $\chi$, but this approximation is not valid on the very large angular scales, which happen to contain the most $f_{\textrm{NL}}$ information. Therefore, we evaluate our theory model using \texttt{Blast}\footnote{\url{https://github.com/sofiachiarenza/Blast.jl}} \citep{chiarenza2024blast, chiarenzainprep}, an algorithm for calculating angular power spectra without employing the Limber approximation or assuming a scale independent growth rate, based on the use of Chebyshev polynomials. 
The code assumes that the power spectrum can be factorized as
\begin{equation}
    P(k) = P_{\textrm{lin}}(k) + (P_\textrm{nl}-P_\textrm{lin})(k)
\end{equation}
The non-linear part is negligible until $\ell > 200$, where the Limber approximation works very well. Therefore, for multipoles $\ell < 200$, we compute the signal using the full non-Limber expression evaluated using $P_\mathrm{lin}(k)$, and then add the non-linear correction $(P_{\mathrm{nl}} - P_{\mathrm{lin}})$ evaluated with Limber approximation. For $\ell > 200$, \texttt{Blast} uses the Limber approximation with $P_\mathrm{nl}(k)$.

The modeling also includes the contributions from lensing magnification and redshift-space distortions. Lensing magnification accounts for the changes on the background density caused by foreground structures. This alters the observed number counts and is effectively correlated with CMB lensing, leading to an extra contribution to the cross-correlation:
\begin{equation}
    C_\ell^{\kappa \mu} = \frac{2}{\pi} \int \mathrm{d}z_1 \frac{W_\kappa(z_1)}{\chi^2(z_1)} \int \mathrm{d}z_2 \frac{W_\mu(z_2)}{\chi^2(z_2)} \int \mathrm{d}k \,P_{mm}(k, z_1, z_2) j_\ell(k\chi(z_1)) j_\ell(k\chi(z_2))
\label{eqn:magbias}
\end{equation}
where the magnification bias kernel $W_\mu(z)$ is
\begin{equation}
   W_\mu(z) =  \frac{3}{2}\frac{H_0^2}{c^2}\Omega_m\,\chi(z)(1+z)\int_z^{+\infty}\mathrm{d}z'n(z')(5s(z')-2)\left(1-\frac{\chi(z)}{\chi(z')}\right)
\end{equation}
The inner integral runs over the distribution of galaxies, and $s \equiv \frac{d\log_{10}{n}}{dm}$ \citep{Scranton05} is the response of the number density $n$ to achromatic changes in the brightness $dm$. In Sec.~\ref{sec:data}, we describe how $s$ is estimated for our sample.
\begin{figure}[h!]
    \centering
    \includegraphics[width=\textwidth]{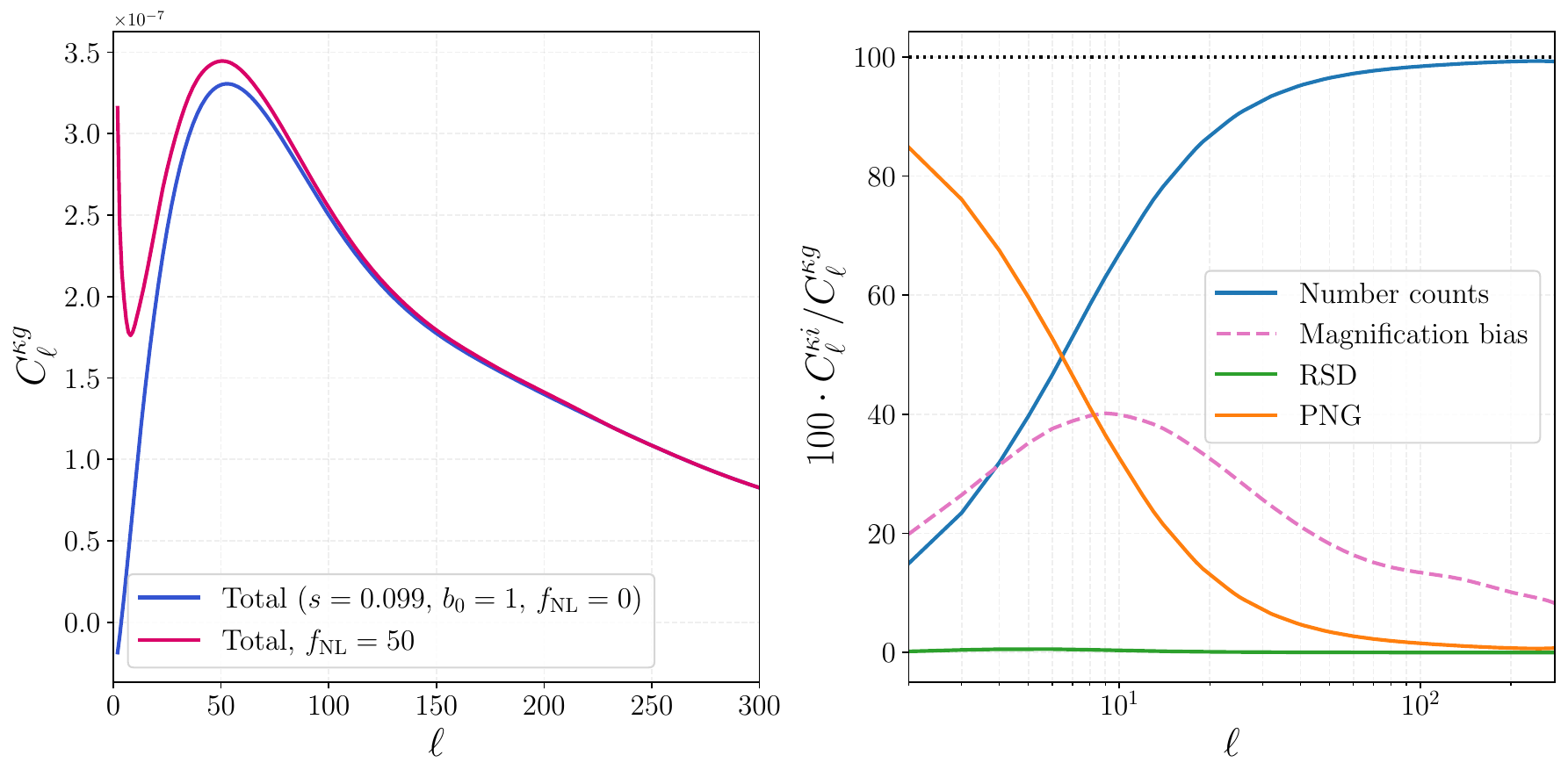}
    \caption{\textit{Left:} Total quasar-CMB lensing cross-correlation in the first tomographic bin ($0.8<z<2.1$). The fiducial model with $f_{\mathrm{NL}}=0$ is presented in blue, and with $f_{\textrm{NL}} = 50$ in magenta. \textit{Right:} Contributions of each term to Eq.~\ref{eqn:total_model} as a fraction of the fiducial model with $f_{\textrm{NL}} = 50$. Negative terms are shown as dashed lines.}
    \label{fig:components_of_clkg}
\end{figure}

Redshift space distortions also contribute to the observed number counts by causing galaxies to be observed in a different redshift shell due to their peculiar velocities:
\begin{equation}
    C_\ell^{\kappa \textrm{RSD}} = \frac{2}{\pi} \int \mathrm{d}z_1 W_{\kappa}(z_1) \int \mathrm{d}z_2 W_{\mathrm{RSD}}(z_2) \int \mathrm{d}k P_{mm}(k, z_1, z_2) j_\ell(k\chi(z_1)) j_\ell''(k\chi(z_2))
\end{equation}
The RSD kernel can be expressed as:
\begin{equation}
    W_{\mathrm{RSD}}(z)=\frac{H(z)}{c}f(z)n(z)
\end{equation}
where $f(z)$ is the logarithmic derivative of the growth rate with respect to scale factor, $f \equiv \frac{d\ln{D}}{d\ln{a}}$, and the spherical Bessel function is replaced with its second derivative.
The only sensitivity to $f_{\mathrm{NL}}$ is through $C_\ell^{\kappa \delta}$, as neither $C_{\ell}^{\kappa \mu}$ nor $C_{\ell}^{\kappa \mathrm{RSD}}$ depend on the galaxy bias.
As discussed, \fnl{} leaves an imprint in the clustering signal through the scale dependent bias, defined in Eq.~\ref{eqn:bias_total}. In the presence of PNG, the number counts kernel, defined in Eq.~\ref{eqn:number_counts}, becomes:
\begin{equation}
    W_{\delta+\mathrm{PNG}}(z) = \frac{H(z)}{c}\left(b_1(z)+\Delta b(k,z)\right)n(z).
\end{equation}
This can be factorized:
\begin{equation}
    W_{\delta+\mathrm{PNG}}(z) = W_\delta(z) + W_{\mathrm{PNG}}(z),
\end{equation}
with
\begin{equation}
    W_\mathrm{PNG} (z)= \frac{H(z)}{c}f_{\mathrm{NL}}b_\Phi(z)n(z).
\end{equation}
Keeping in mind that \texttt{Blast} performs the inner $k$-integral first, and that this step is affected by the presence of the transfer function in the scale-dependent bias, it can be worked out that the PNG contribution to the cross-correlation is:
\begin{equation}\label{eqn:fnl_term}
C_\ell^{\kappa \mathrm{PNG}} = \frac{2}{\pi}
\int \mathrm{d}z_1\, \frac{W_\kappa(z_1)}{\chi^2(z_1)}
\int \mathrm{d}z_2\, W_{\mathrm{PNG}}(z_2)
\int \mathrm{d}k\,
P_\Phi(k)\,
T_{\Phi\rightarrow\delta}(k,z_1)\,
j_\ell(k\chi_1)\, j_\ell(k\chi_2),
\end{equation}
where the matter power spectrum has been written as $P_{mm}(k,z_1,z_2)=P_\Phi(k)\,T_{\Phi\rightarrow\delta}(k,z_1)\,T_{\Phi\rightarrow\delta}(k,z_2)$.
The reason why Eq.~\ref{eqn:fnl_term} contains only one transfer function is that the scale-dependent bias scales as
$\Delta b(k,z)\propto 1/T_{\Phi\rightarrow\delta}(k,z)$, 
which cancels the second transfer-function factor from $P_{mm}$.
Effectively, we can write our final model as:
\begin{equation}
    C_{\ell}^{\kappa g} = C_{\ell}^{\kappa \delta} + C_{\ell}^{\kappa \mu} + C_{\ell}^{\kappa \textrm{RSD}} + C_{\ell}^{\kappa \textrm{PNG}}
\label{eqn:total_model}
\end{equation}
with \(C_{\ell}^{\kappa \textrm{PNG}} \propto f_{\mathrm{NL}}\). For a more technical description of the algorithm, see \citep{chiarenzainprep}.

Fig.~\ref{fig:components_of_clkg} shows the fractional contributions of magnification, RSD and PNG to the total $C_\ell^{\kappa g}$ signal with  $f_{\textrm{NL}} = 50$ in the first tomographic bin ($0.8<z<2.1)$. The fiducial values of $b_0^i = 1$ and $s = 0.099$ are assumed.
The magnification term is a considerable fraction ($\sim 15\%$) of the clustering term, and rises in a scale-dependent way that is approximately degenerate with $f_{\textrm{NL}}$ at $\ell > 30$, emphasizing the importance of low multipoles. 
Despite this degeneracy, the magnification bias slope is measured sufficiently accurately that it does not worsen the $f_{\textrm{NL}}$ constraints. The RSD term is very subdominant, only rising to 1\% of the fiducial model at $\ell < 6$. Finally, the $f_{\textrm{NL}}$ contribution dominates for $\ell<10$ and constitutes a non negligible part of the signal up to $\ell \approx 100$. Although $f_{\textrm{NL}}=50$ is a very high and unrealistic value for this parameter, it shows how a non-Limber evaluation of the theory model is crucial to probe the PNG signal. 
\begin{figure}[h!]
    \centering
    \begin{subfigure}[b]{0.3\textwidth}
        \centering
        \includegraphics[width=\textwidth]{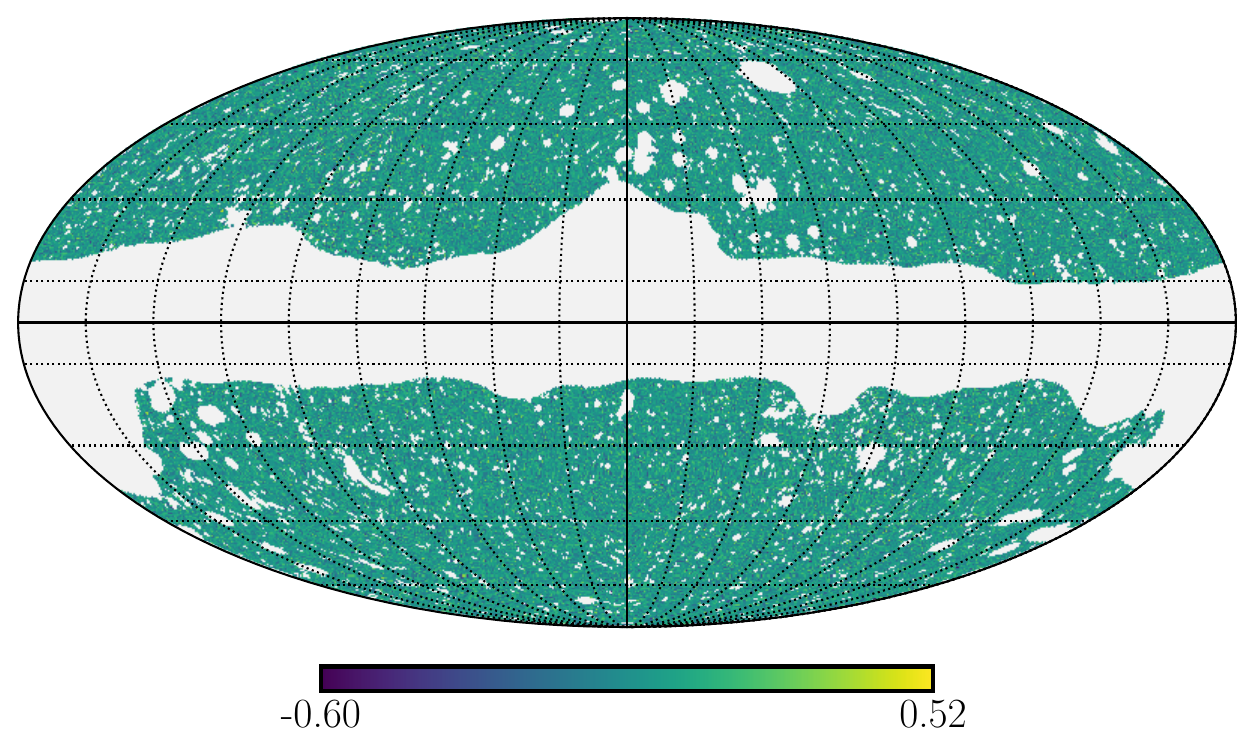}
        \caption{\Planck PR4 lensing map.}
        \label{fig:fig1}
    \end{subfigure}
    \hfill
    \begin{subfigure}[b]{0.3\textwidth}
        \centering
        \includegraphics[width=\textwidth]{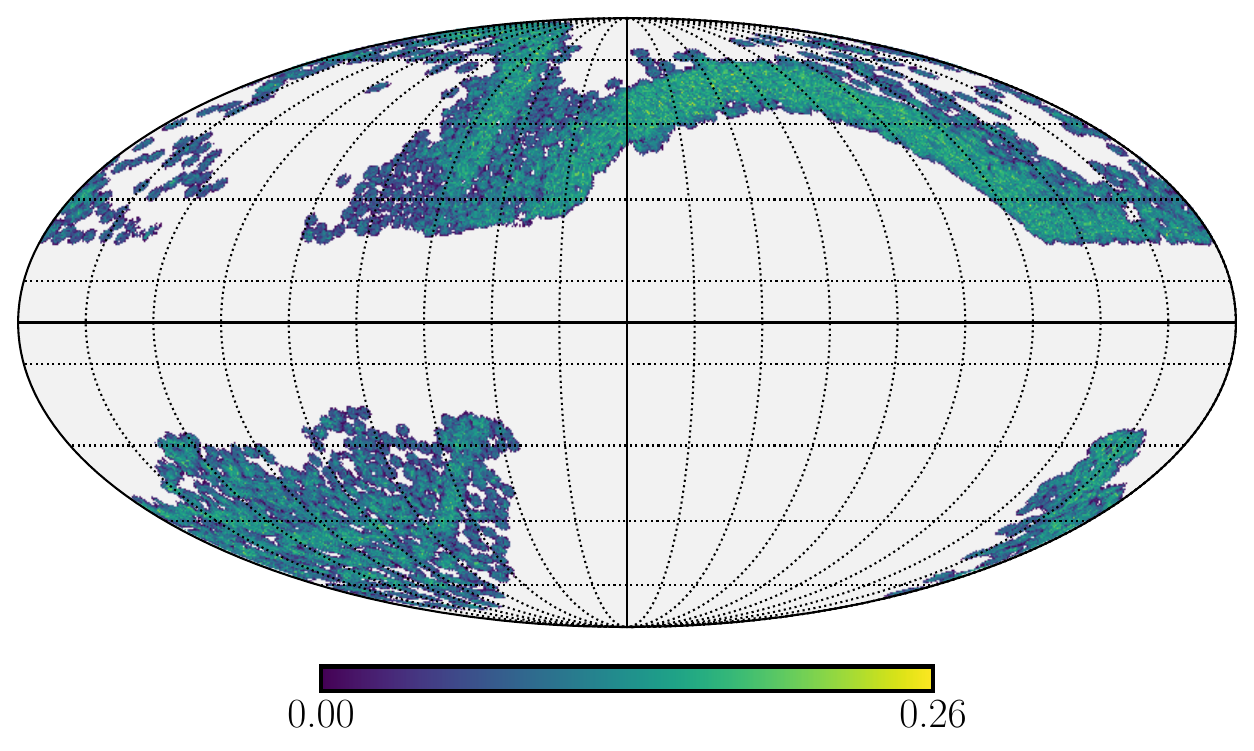}
        \caption{DESI DR1 quasar sample.}
        \label{fig:fig2}
    \end{subfigure}
    \hfill
    \begin{subfigure}[b]{0.3\textwidth}
        \centering
        \includegraphics[width=\textwidth]{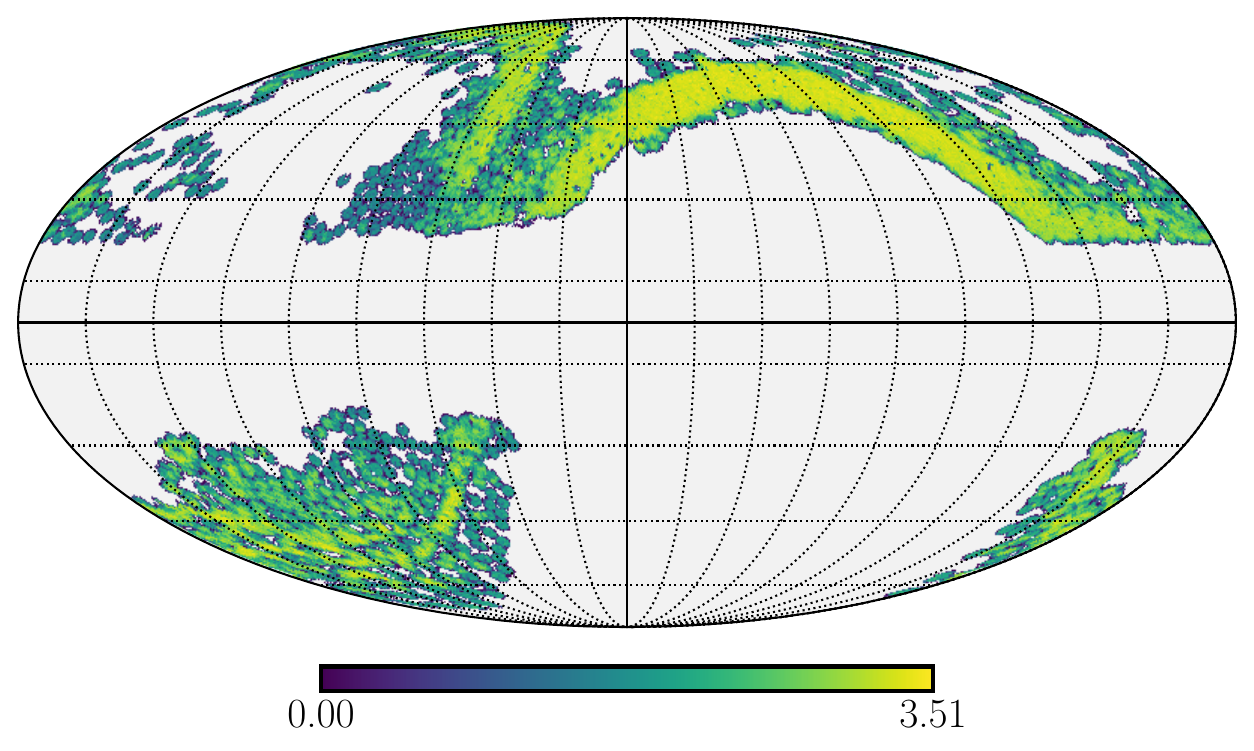}
        \caption{Completeness mask.}
        \label{fig:fig3}
    \end{subfigure}
    \caption{Data used for the cross-correlation (same as in Ref.~\cite{de2025cosmology}). Panel (a) shows the \Planck PR4 lensing convergence map, $\kappa$, together with its smoothed mask, obtained by applying a Gaussian filter with $1^{\circ}$ FWHM. Panel (b) displays the DESI DR1 quasar number counts for the full sample spanning the redshift interval $0.8 \leq z \leq 3.5$, while the corresponding completeness mask is shown in panel (c). For visualization, we adopt a \texttt{HEALPix} resolution of $N_{\rm side}=128$, whereas all computations are performed at $N_{\rm side}=2048$. All maps are displayed in a Mollweide projection and are presented in the Galactic coordinate system.}
    \label{fig:data}
\end{figure}

\section{Data}
\label{sec:data}
We cross-correlate quasars from the DESI survey with lensing mass maps obtained by the \Planck satellite, namely the PR4 convergence maps \citep{akrami2020planck}. The data are visualized in Fig.~\ref{fig:data}. In Sec.~\ref{sec:qso_sample} we summarize the quasar sample and in Sec.~\ref{sec:cmb_lensing} we briefly present the employed CMB lensing data. 
\begin{figure}[h!]
    \centering
    \includegraphics[width=\linewidth]{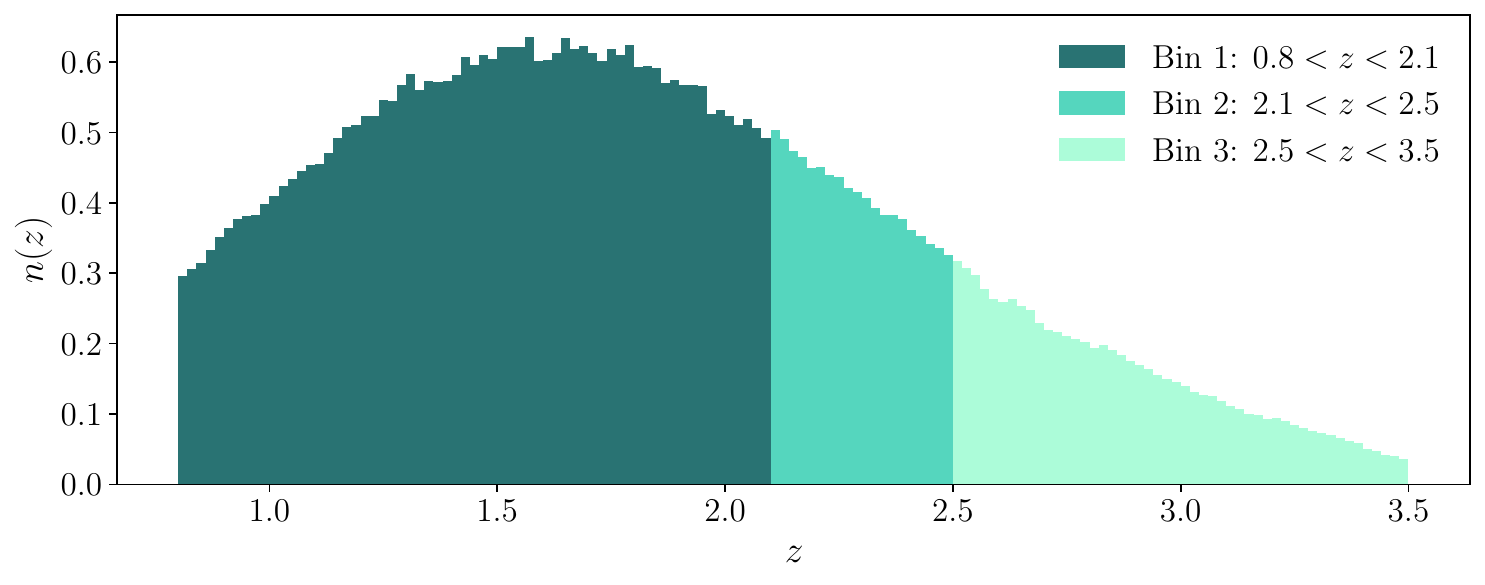}
    \caption{Normalized redshift distribution $n(z)$ of the DESI DR1 spectroscopic quasar sample. The different colors identify the three non-overlapping redshift bins.}
    \label{fig:DESI_DR1_dNdz_hist}
\end{figure}
\subsection{DESI DR1 Spectroscopically confirmed quasars}
\label{sec:qso_sample}
Our analysis employs a spectroscopic catalogue of $1\,223\,391$ quasars in the redshift interval $0.8\le z\le 3.5$ taken from the DESI DR1 quasar sample \citep{chaussidon2023target} released with the survey’s first public data release \citep{abdul2025data}. DESI is a highly multiplexed fiber-fed spectrograph located on the 4m Mayall telescope at the Kitt Peak National Observatory \citep{DESI2016b, abareshi2022overview}. It has the capability to record up to $5\,000$ spectra in a single exposure \citep{aghamousa2016desi,silber2022robotic,miller2023optical,poppett2024overview,schlafly2023survey}.  
\begin{table}[h!]
    \centering
    \setlength{\tabcolsep}{10pt}
    \begin{tabular}{cccccccc}
    \hline\hline \vspace{-2ex} \\ 
    Label & $z$-range & $N_{\rm qso}$ & Shot Noise & $\overline{z}$ & $z_{\rm eff}$ & $s_{\mu}$ & $f_{\rm sky}^{\rm PR4\,\times\,DR1}$ \\ [0.5ex]
    \hline \vspace{-2ex} \\ 
    $g_1$ & $0.8 \le z < 2.1$ & 856\,831 & $\phantom{1}2.6\times10^{-6}$ & 1.49 & 1.44 & 0.099 & 19.8 \\ [0.5ex]
    $g_2$ & $2.1 \le z < 2.5$ & 194\,754 & $11.2\times10^{-6}$ & 2.28 & 2.27 & 0.185 & 18.9 \\ [0.5ex]
    $g_3$ & $2.5 \le z \le 3.5$ & 171\,806 & $12.7\times10^{-6}$ & 2.85 & 2.75 & 0.244 & 18.6 \\ [0.5ex]
    \hline
    \end{tabular}
    \caption{Overview of DESI DR1 quasar samples used for the cross-correlation with \Planck{} CMB lensing.  We list the three redshift bins $\{g_1,g_2,g_3\}$, their redshift ranges, quasar counts, shot-noise levels, mean redshift $\overline{z}$, effective redshift $z_{\rm eff}$ \citep{Sailer2024}, magnification bias $s_{\mu}$, and the sky-fraction overlap with the PR4 CMB-lensing convergence maps, $f_{\rm sky}^{\rm PR4\times DR1}$ (in percent).}
    \label{tab:DESI_DR1_data}
\end{table}
Throughout a eight-year campaign, DESI aims to map up to $17\,000\,\mathrm{deg}^2$, which will provide precise measurements of baryon acoustic oscillations (BAO) and redshift-space distortions (RSD) from both galaxies and quasars across $0\lesssim z\lesssim 3.5$ \citep{levi2013desi,aghamousa2016desi}.
Following successful validation of the survey \citep{adame2024validation} and an early data release \citep{adame2024early}, the data from the first year has already provided competitive BAO constraints \citep{adame2025bao,adame2025desilya,adame2025desiqso}, accurate RSD measurements \citep{adame2024desi}, and the currently strongest large-scale structure limit on primordial non-Gaussianity, derived from the quasar power spectrum \citep{chaussidon2024}.

Table~\ref{tab:DESI_DR1_data} summarizes the DR1 quasar subsamples, while Fig.~\ref{fig:DESI_DR1_dNdz_hist} illustrates their redshift distributions alongside the tomographic binning scheme used in this analysis. This scheme is consistent with the accompanying study of DESI DR1 quasars provided in \citep{de2025cosmology}. The sample is divided into three redshift bins to retain radial information. The first bin corresponds to the DESI fiducial range, $0.8<z<2.1$, while the remaining quasars, extending up to $z=3.5$, are split into two approximately equally populated bins, $2.1<z<2.5$ and $2.5<z<3.5$.
We utilize the clustering catalogues described in \citep{adame2025desisample,ross2025construction} together with redshifts produced by the DESI spectroscopic pipeline \citep{guy2023spectroscopic,bailey2024redrock,brodzeller2023performance,anand2024archetype}.

Table~\ref{tab:DESI_DR1_data} also reports the measured values of magnification bias for the three tomographic bins. We measure the magnification bias slope, $s \equiv d\log_{10}N/dm$, by perturbing the Legacy Imaging Survey DR9 photometry \citep{Dey19} uniformly by $\pm 0.05$ mag, re-running quasar target selection, and measuring the resulting change in number density. The values obtained from brightening and dimming are consistent, so we adopt their mean as our fiducial slope. Since quasars are dominated by point sources and selected using total magnitudes rather than fiber magnitudes, it is an excellent approximation to treat lensing magnification as a simple brightening or dimming of their flux. This procedure measures the slope using the observed (post-magnification) magnitude distribution, but the induced bias is negligible, as magnification only scatters a small fraction of quasars across the flux limit. We find that the slope varies with redshift, and therefore use the value appropriate for each bin: $s = \,0.099$ for $0.8<z<2.1$, $s = \,0.185$ for $2.1<z<2.5$, and $s = \,0.244$ for $2.5<z<3.5$.

\subsection{\Planck PR4 CMB lensing map}
\label{sec:cmb_lensing}
Gravitational lensing by large–scale structure deflects CMB photons, imprinting subtle distortions in the temperature and polarization fields that can be exploited to reconstruct the lensing potential \citep[see, e.g.,][for a review]{lewis2006weak}. For this work we employ the \Planck PR4 CMB lensing convergence maps, $\kappa$, which are publicly available\footnote{\url{https://pla.esac.esa.int} and \url{https://github.com/carronj/planck_PR4_lensing}.} \citep{akrami2020planck}. Starting from the global minimum–variance (GMV) spherical-harmonic coefficients $\kappa_{\ell m}$, we resample the data onto a \texttt{HEALPix} grid with \texttt{NSIDE}=2048 \citep{gorski2005healpix}.

The PR4 release benefits from the \textsc{NPIPE} processing pipeline \citep{akrami2020planck}, which provides improved, uniformly processed CMB maps that serve as inputs to the lensing reconstruction. The reconstruction itself was upgraded in PR4: the collaboration adopted a GMV estimator that incorporates temperature–polarization correlations \citep{maniyar2021quadratic} and implemented an anisotropic filtering and local-noise weighting scheme \citep{carron2022cmb}. Combined with the $\sim8\%$ additional CMB observations obtained during satellite repointing, these improvements yield an overall $\sim20\%$ increase in signal–to–noise. 
As a result, the PR4 lensing map remains signal–dominated up to multipoles $\ell\simeq70$, compared with $\ell\simeq40$ for PR3.
The effective reconstruction noise power spectrum, $N_{\ell}^{\kappa\kappa}$, supplied by the \Planck collaboration, is incorporated in the analytical covariance matrix, described in Sec.~\ref{sec:ps}. The lensing mask retains a sky fraction of $f_{\rm sky}=67.1\%$\footnote{The sky fraction is defined as $f_{\rm sky}=\sum_{i=1}^{N_{\rm pix}}x_i/N_{\rm pix}^{\rm tot}$, where $N_{\rm pix}^{\rm tot}=12\,N_{\rm side}^2\approx5\times10^7$ and $x_i$ is the value of the mask in pixel $i$.}. Its overlap with the tomographic DESI DR1 quasar catalogue is summarized in Table~\ref{tab:DESI_DR1_data}.

\section{Measuring Angular Power Spectra}
\label{sec:ps}
We measure clustering directly from the un-pixelised DESI quasar catalogue using the catalog-based \emph{pseudo}-$C_\ell$ estimator \citep{wolz2025catalog, alonso2019unified}. Our analysis utilizes the \texttt{NaMaster} implementation\footnote{\url{https://github.com/LSSTDESC/NaMaster}} \citep{alonso2019unified}, which builds upon the algorithm described in \citep{lizancos2024harmonic} and leverages the \texttt{DUCC} library\footnote{\url{https://gitlab.mpcdf.mpg.de/mtr/ducc}} for maximal efficiency \citep[e.g.,][]{reinecke2023improved}.  
In the standard pixelized approach, galaxies and randoms are first binned onto a high–resolution \texttt{HEALPix} grid, and the overdensity field is formed as \(1 + \delta_\mathrm{gal}\) by dividing the two maps. While conceptually straightforward, this division can lead to numerical instabilities in pixels where the survey completeness is very low. In early tests with this pixel-based procedure, we observed a significant excess of power on large scales in the measured angular power spectra, which could not be accounted for by known imaging systematic effects. By carefully tracking the origin of this excess, we found it was caused by instabilities in the overdensity maps calculation in low-completeness pixels. Switching to the catalog-based \emph{pseudo}–$C_\ell$ estimator completely resolved this issue, yielding robust measurements across all scales. We analyze angular scales in the range \(4 \le \ell \le 300\). The binning scheme uses \(\Delta\ell = 10\) for \(\ell \le 100\), where most of the signal resides, and \(\Delta\ell = 50\) for \(\ell \ge 100\), yielding $14$ bins in total. Finally, we stress that the \texttt{NaMaster} estimator automatically subtracts the Poisson shot noise from the measured auto–power spectra. We have verified that the shot–noise levels estimated internally by the code agree with the theoretical expectations obtained from the number density in each redshift bin (Table~\ref{tab:DESI_DR1_data}).

Furthermore, the pixelized approach is subject to aliasing and pixel–window function effects. Mitigating these problems usually requires ultra-fine grids (with an arbitrary choice of \texttt{NSIDE}) and very large random catalogues, driving up both memory and computational costs. 
\begin{figure}[h!]
\centering
    \includegraphics[width=0.6\linewidth]{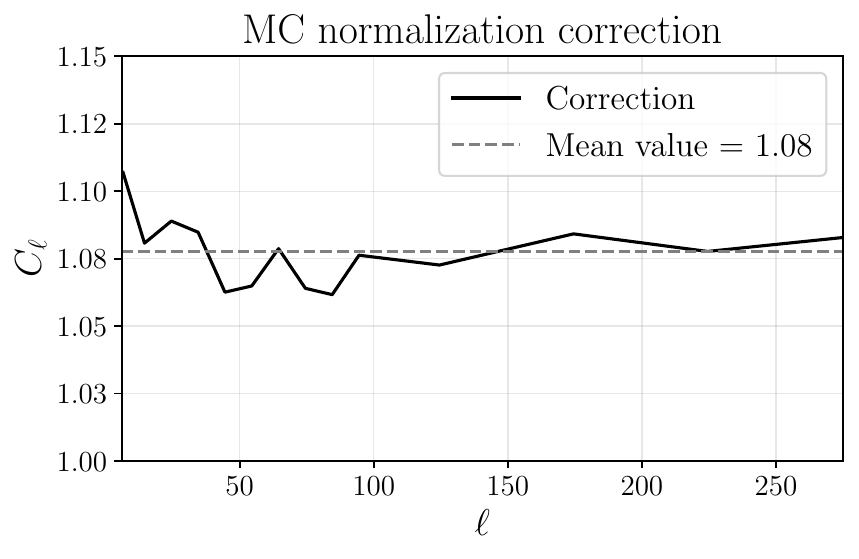}
    \caption{Monte Carlo normalization correction applied to the cross-spectra $C_\ell^{\kappa g_i}$ which amounts to an approximately constant amplitude change of $\approx 8\%$.}
    \label{fig:mcnorm}
\end{figure}
The catalog-based pipeline instead adopts the ``FKP'' approach  from three-dimensional analyses by working with the linear difference \(n_{\mathrm g} - \alpha\,n_{\mathrm r}\) instead of the ratio of data over randoms catalogs.  Because this operation is linear, we can simply subtract the direct spherical–harmonic transforms of data and scaled randoms, avoiding numerical instabilities and pixel-related problems. In the limit that the objects are point-like, the spherical harmonic transform of these fields is simply a sum of spherical harmonics evaluated at the positions of the objects, \(Y_{\ell m}(\theta_i, \phi_i)\), making the evaluation of the estimator straightforward. For a detailed description of the algorithms, the reader is referred to \citep{wolz2025catalog, lizancos2024harmonic}.

Pixel-free estimators offer a promising approach for future analyses involving cross-correlations with discrete tracers, however the corresponding analytic covariance calculations for catalog-based fields are still under development \citep{tessore2025shotnoiseclusteringpower}. Therefore, following the approach used, for example, in \citep{fabbian2025, maus2025joint}, we adopt the standard analytic Gaussian, window-convolved covariance approximation implemented in the \texttt{gaussian\_covariance} method of the \texttt{NaMaster} package to estimate the errors in our analysis. Since this covariance routine requires a pixelized map of the data field, we construct such a map from the data $a_{\ell m}$'s obtained with the catalog-based estimator, generating \texttt{HEALPix} maps at \texttt{NSIDE}~$=2048$.
Moreover, this approach also requires some theory spectra, which, as discussed in Sec.~\ref{sec:theory}, we generate using \texttt{Blast} \citep{chiarenza2024blast, chiarenzainprep}. The code gives predictions for \(C_\ell^{\kappa g_i}\) and \(C_\ell^{g_ig_j}\) (where we add the shot noise contribution as in Table~\ref{tab:DESI_DR1_data} to \(C_\ell^{g_ig_i}\)) in the fiducial cosmology. For the lensing power spectrum \(C_\ell^{\kappa \kappa}\) we use the publicly available PR4 lensing spectra with the provided lensing reconstruction noise \(N_{\kappa\kappa}\). As in \citep{de2025cosmology},  we use an iterative approach: we fit the cosmological and nuisance parameters with the covariance matrix evaluated with the theory curves just described, and use the resulting best-fit parameters to recalculate the fiducial spectra and tune the covariance matrices. 

\subsection{Monte Carlo normalization correction for the \Planck lensing maps}
A well-known effect in CMB lensing reconstruction is a misnormalization of the reconstructed field caused by the masks and anisotropic filtering applied in order to perform the reconstruction. The mode coupling introduced in this procedure is not modeled into the cross-spectra obtained with the \texttt{NaMaster} algorithm. As a result, the reconstructed convergence $\hat{\kappa}$ does not have the same normalization as the true convergence $\kappa$, and a correction is required \citep{farren2024atacama, benoit2013full, carron2023real}. 

In practice, this correction is determined through simulations: the basic idea is to compare the cross-correlation between the input convergence (appropriately masked) and the reconstruction with the known input power spectrum. This yields a Monte Carlo normalization factor that is dependent on the mask and the footprint: 
\begin{equation}
    A_{\ell}^{\mathrm{MC}} = 
    \frac{C_{\ell}^{\kappa_{\mathrm{in}, \kappa-\mathrm{mask}} \, \kappa_{\mathrm{in}, g-\mathrm{mask}}}}
         {C_{\ell}^{\hat{\kappa} \, \kappa_{\mathrm{in}, g-\mathrm{mask}}}},
\end{equation}
where $\hat{\kappa}$ is the masked CMB lensing reconstruction, $\kappa_{\mathrm{in}, \kappa-\mathrm{mask}}$ is the input lensing convergence masked with the lensing mask, and $\kappa_{\mathrm{in}, g-\mathrm{mask}}$ is the input convergence masked using the galaxy mask. An unbiased estimate of the CMB lensing cross-spectrum is then obtained as
\begin{equation}
    C_\ell^{\hat{\kappa}_{\mathrm{MC}} g} = C_\ell^{\hat{\kappa} g} \, A_{\ell}^{\mathrm{MC}}.
\end{equation}
Following \citep{de2025cosmology}, we implement this correction in a simulation-based, mode-by-mode fashion. Specifically, we generate a suite of CMB lensing reconstructions using the appropriate mask:
\begin{equation}
    T_L \equiv \frac{\sum_{\ell} W_{L \ell} \sum_i^{N_{\mathrm{sim}}} \sum_{m=-\ell}^{\ell}\left\{M^\kappa \kappa^i\right\}_{\ell m}\left\{M^q \kappa^i\right\}_{\ell m}^*}{\sum_{\ell} W_{L \ell} \sum_i^{N_{\mathrm{sim}}} \sum_{m=-\ell}^{\ell}\left\{M^\kappa \hat{\kappa}^i\right\}_{\ell m}\left\{M^q \kappa^i\right\}_{\ell m}^*},
\end{equation}
where $W_\mathrm{L}$ indicates the bandpower window functions, $M$ denotes the mask, $\kappa$ the input (and $\hat{\kappa}$ the reconstructed) convergence, $i$ indexes the simulations, and ${XY}_{\ell m} \equiv \int \mathrm{d}^2 \hat{\mathbf{n}} Y_{\ell m}^*(\hat{\mathbf{n}}) X(\hat{\mathbf{n}}) Y(\hat{\mathbf{n}})$ with $Y_{\ell m}$ the spherical harmonics. For the \Planck PR4 maps over the DESI DR1 footprint, this correction is at the level of $\approx 8\%$ on the scales of interest (see Fig.~\ref{fig:mcnorm}).

\section{Imaging systematic weights validation}
\label{sec:sys_weights}
Spatial variations in imaging quality and foregrounds imprint spurious modulations in the angular number density of DESI targets. If left uncorrected, these modulations contaminate the largest‐scale Fourier modes that are most sensitive to primordial non-Gaussianity. 

\subsection{The DESI systematic weights}
DESI adopts the template-fitting approach introduced in recent large-scale structure surveys. The technique derives a per-object weight, \(w_{\mathrm{sys}}\), by quantifying how the target surface density varies with each imaging attribute and then correcting for those dependencies.
More specifically, DESI attaches three multiplicative weights to every object:
\begin{equation}\label{eqn:galaxy_weights}
    w_{\mathrm{tot}} = w_{\mathrm{comp}}\,w_{\mathrm{sys}}\,w_{\mathrm{zfail}},
\end{equation}
ensuring that the ratio of weighted data to weighted randoms remains flat across the footprint once all known observational selection effects are removed. Completeness weights \(w_\mathrm{comp}\) correct for the probability that a target is allocated a fiber, and redshift failure weights, \(w_\mathrm{zfail}\), mitigate spatial fluctuations in redshift success produced by variations in exposure time, focal-plane position, and hardware status. The imaging systematics weights $w_\mathrm{sys}$, the focus of this section, remove spurious target-density fluctuations that track imaging conditions such as depth, seeing, dust, and stellar density. Full implementation details for the first two weights are given in \citep{ross2025construction}.

Below, we discuss only the imaging systematic weights, since this is the element we test extensively for its impact on the quasar angular power spectrum. The other two weights, completeness and redshift-failure, are not a concern for our analysis. Systematic weights have many possible choices, including different templates and implementations, and can have a large impact on the results. By contrast, it is well understood how to correctly account for completeness. Redshift-failure weights have a much smaller impact than the systematic errors and therefore do not significantly affect the analysis \citep{ross2025construction, ross2020completed}. The DESI catalog in the latest released version (\texttt{v1.5}), comes with three possible sets of systematic weights: the neural net method \texttt{sysnet}, denoted \texttt{WEIGHT\_SN} in the catalogs; the random forest method \texttt{regressis}, denoted \texttt{WEIGHT\_RF}; the linear method applied to eBOSS \citep{ross2020completed} LSS catalogs, identified as \texttt{WEIGHT\_IMLIN}.
In the present work, we only consider two different sets of linear weights as it is known that both machine learning based weights, namely the \texttt{sysnet} neural network \citep{rezaie2020improving} and the \texttt{regressis}\footnote{\url{https://github.com/echaussidon/regressis}} random forest weights \citep{Chaussidon22}, over-correct for excess large scale power introduced by systematics, especially in the case of the quasar sample \citep{chaussidon2024, ross2025construction, Chaussidon22}. The second set of linear weights is also computed with \texttt{regressis} at \texttt{HEALPix} map level using a linear regression algorithm. Note that, compared to the catalog linear weights, the linear regression here does not fit the data to binned statistics but rather the fluctuation at \texttt{HEALPix} map level. The first are identified as \texttt{WEIGHT\_IMLIN}, the latter as \texttt{WEIGHT\_Linear}.

\subsection{Correcting for excess mode removal}
As described above, we test two sets of linear systematic weights that are derived from the same imaging templates but differ in their implementation. Accurately assessing the performance of such weights is crucial, as they can over-correct and remove large-scale clustering signal, thereby weakening the \fnl{} constraints. The correction for this mode removal by reweighting modes is related to the angular integral constraint (AIC) correction \cite{adame2025desisample}.
This effect has been shown to be particularly severe for neural network–based weighting schemes \citep{chaussidon2024}, which we therefore do not consider in this analysis. Nevertheless, it remains important to quantify any residual bias for the two linear weighting schemes tested here. To quantify the impact of our systematic–mitigation weighting scheme, we use custom-generated mocks with no imaging systematic contamination, so that tracer densities are, on average, uncorrelated with the imaging features. Any deviation in the angular power spectra after applying the systematic weights, will be due to the mitigation method itself and must be corrected on the measurement on the data.  

\subsubsection{Creating the mocks}
We have created $500$ uncontaminated, catalog-based lognormal mocks that accurately reproduce the DESI DR1 footprint and clustering properties. This number of realizations is sufficient to ensure that the uncertainty associated with the mock-based correction remains well below the statistical errors across the relevant scales. The first step is to generate correlated Gaussian maps for the three redshift bins used in the analysis ($0.8<z<2.1$, $2.1<z<2.5$, $2.5<z<3.5$) and for the CMB lensing map. This is done using the \texttt{healpy} package \citep{zonca2019healpy}, specifically the \texttt{synfast} function, which synthesizes maps at a chosen resolution from input angular power spectra $C_\ell$.
The theoretical $C_\ell$ are computed with the \texttt{GLASS}\footnote{\url{https://github.com/glass-dev/glass}} package \citep{tessore2023glass}, which allows the generation of lognormal fields with prescribed two-point statistics. Because the lognormal transformation, defined as:
\begin{equation}
f(X;\lambda) = \lambda \left( e^X - 1 \right),
\end{equation}
is nonlinear and modifies the underlying statistics of the field, \texttt{GLASS} determines, through an iterative algorithm, the Gaussian input power spectrum $G_\ell$ that will yield the target $C_\ell$ after applying the transformation. In our case, we set the shift-parameter $\lambda$ to $1$ and transform the three Gaussian fields representing the tomographic bins. The target angular spectra $C_\ell$ are computed with \texttt{Blast} \citep{chiarenza2024blast, chiarenzainprep} and the bias is adjusted to match the data, ensuring that the final mocks reproduce the same large-scale clustering statistics as the data.
\begin{figure}[h!]
    \centering
    \begin{subfigure}{0.49\textwidth}
        \centering
        \includegraphics[width=\linewidth]{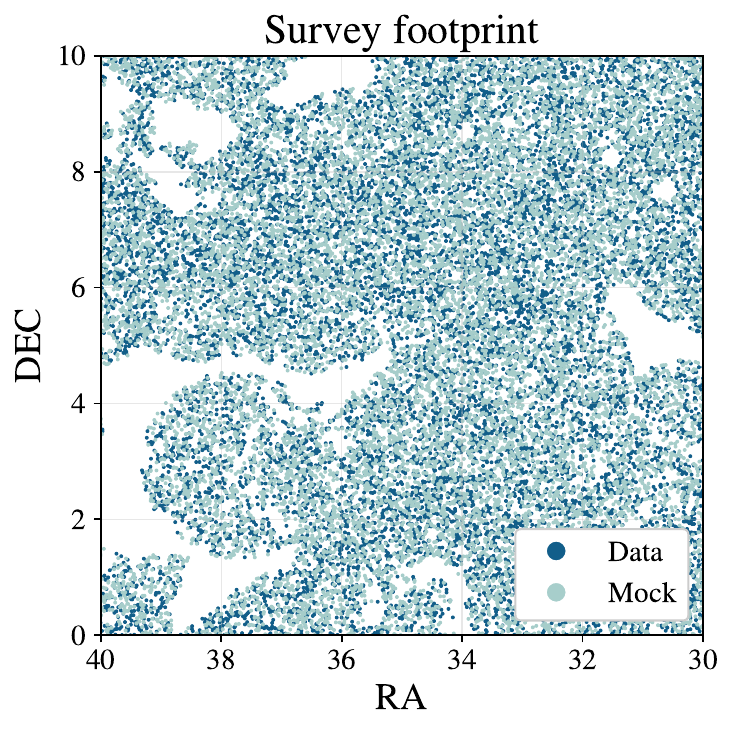}
    \end{subfigure}
    \begin{subfigure}{0.49\textwidth}
        \centering
        \includegraphics[width=\linewidth]{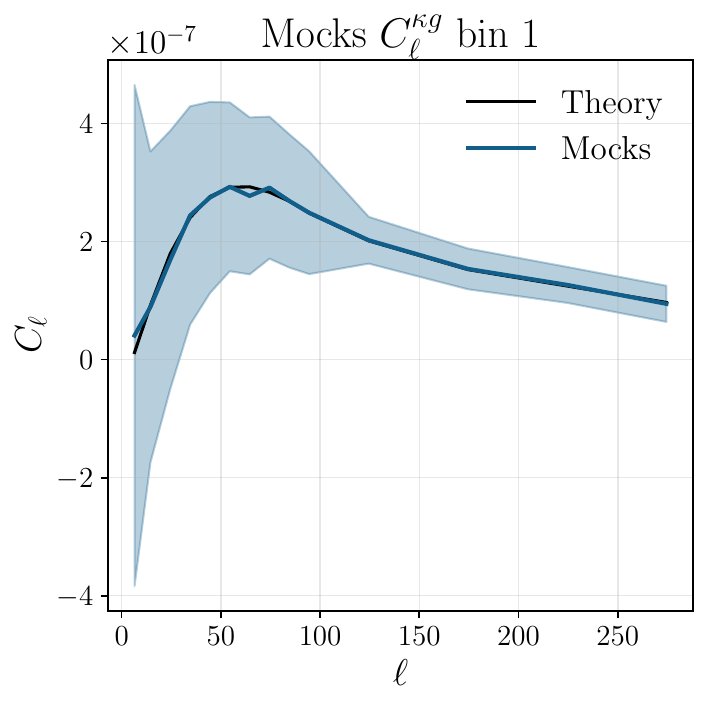}
    \end{subfigure}
    \caption{\textit{Left}: A zoom into the survey footprint. In darker blue, the DESI DR1 quasars, while in light blue are the galaxies making up one of the $500$ mocks. The agreement between the two footprint is very good, confirming the success of our procedure to replicate the DESI survey geometry. \textit{Right}: Average angular power spectrum $C_\ell^{\kappa g}$ over $500$ mocks with the corresponding $1\sigma$ error bars. As desired, the mocks have a power spectrum that matches the input theory curve.}
    \label{fig:mock_check}
\end{figure}
We convert the pixelized lognormal maps into mock catalogs by Poisson-sampling the number of galaxies in each pixel according to its overdensity. Each mock galaxy is then assigned sky coordinates $(\mathrm{RA}, \mathrm{DEC})$ by randomly perturbing the position around the pixel center using an effective pixel radius $\theta_\mathrm{pix}=\sqrt{A_\mathrm{pix}}/2$. During this process, the maps are normalized so that the resulting mock catalogs contain approximately the same number of objects per redshift bin as the data. As this process is very approximate on small scales, we generate the maps at a very high resolution (\texttt{NSIDE} $= 8192$) to ensure that, on the relevant scales ($\ell<300$), the resulting angular clustering would not be affected by this sampling procedure. 

The final step to obtain realistic mocks is to apply angular cuts and downsampling to reproduce the survey footprint and completeness. The DESI DR1 footprint exhibits substantial small-scale structure, as the survey completeness remains low at this stage. To reproduce these variations as accurately as possible, we perform four successive downsampling steps, designed to account for the main observational effects impacting the real survey, while avoiding the computational cost of running each mock through the \texttt{fiberassign} software \citep{bianchi2025characterization} and detailed veto masks. In particular, we select areas of the sky covered during DR1, remove areas around bright stars and regions affected by bad imaging systematic conditions. Finally, we account for the survey completeness by constructing completeness maps directly from existing survey random catalogs. Specifically, we use the official DESI DR1 quasar random files, which encode the spatial completeness pattern of the survey, together with the Legacy Survey DR9 randoms \citep{Dey19}, which are uniform over the sky. Both sets of randoms are pixelized at \texttt{NSIDE} $=2048$, and the ratio of the two maps provides an estimate of the angular completeness. Each mock catalog is then downsampled according to this completeness map, ensuring that the mocks reproduce the same spatial variations in target completeness as the data. This procedure is correct up to the resolution of the maps, which is enough for our analysis, restricted to scales $\ell < 300$.
The results of these footprint and clustering validations are shown in Fig.~\ref{fig:mock_check}. The left panel demonstrates that the mocks accurately reproduce the DESI DR1 survey geometry, as evidenced by their excellent agreement with the official quasar randoms. This confirms that the completeness correction procedure described above effectively transfers the angular selection function to the mock catalogs. The right panel shows that the average cross-correlation $C_\ell^{\kappa g}$ measured from 500 mocks is consistent with the input theoretical prediction, within the expected statistical uncertainties. Together, these tests verify that our catalog-based lognormal mocks correctly reproduce both the survey footprint and the target clustering statistics.
\begin{figure}[h!]
    \includegraphics[width=\linewidth]{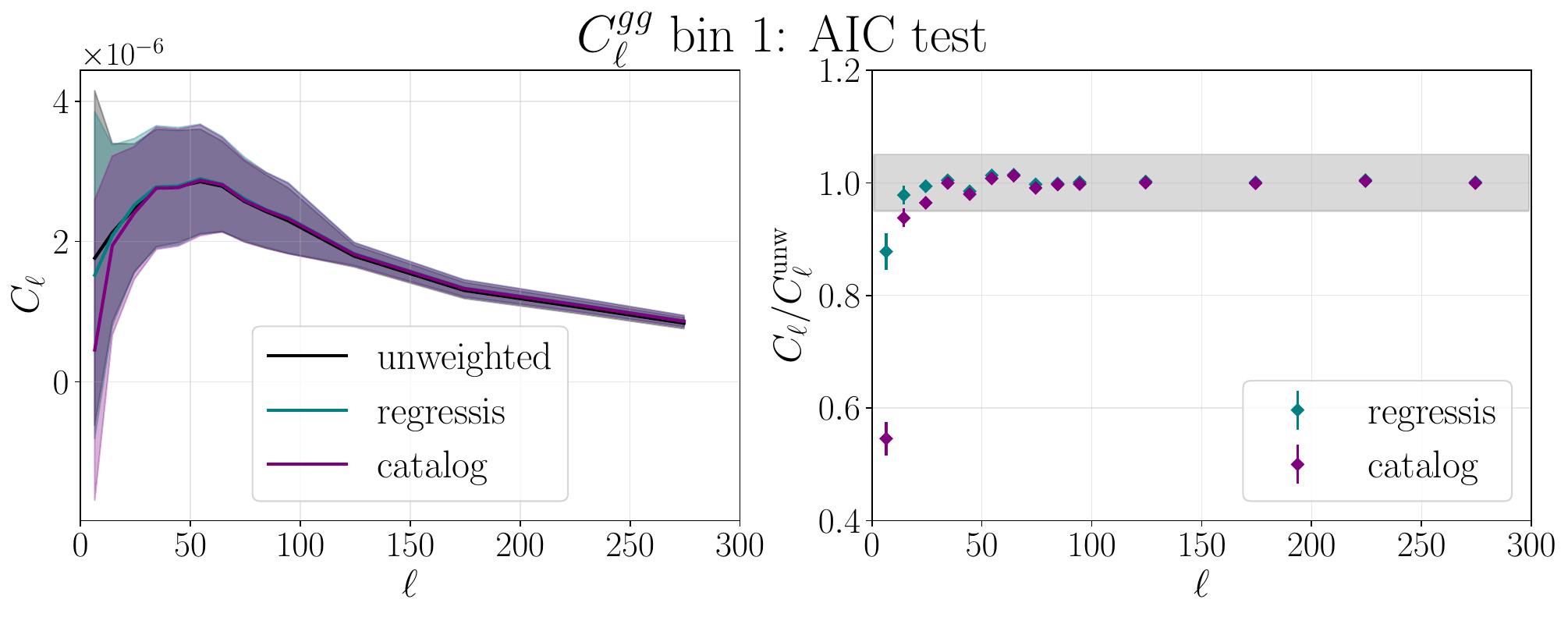}
    \caption{Left: Average angular auto power spectrum $C_\ell^{gg}$ of $500$ mocks with different linear  weighting schemes, in the first redshift bin considered in the analysis ($0.8<z<2.1$). Right: Ratio of the weighted to unweighted power spectra. The ratio deviates from $1$ on large scales, showing that the weights are removing signal from the analysis. The shaded grey area highlights the 1\% region around unity. }
    \label{fig:aic_gg}
\end{figure}
\begin{figure}[h!]
    \includegraphics[width=\linewidth]{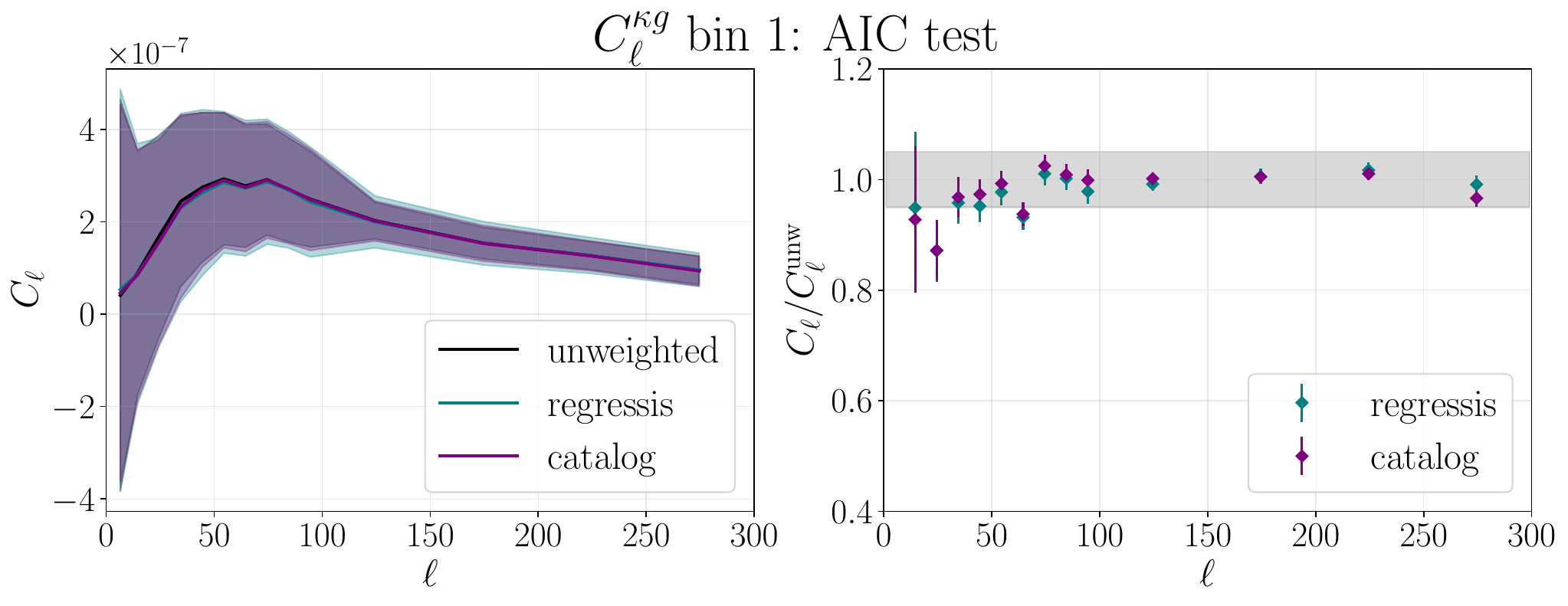}
    \caption{Left: Average angular cross-power spectrum $C_\ell^{\kappa g}$ of $500$ mocks with different linear weighting schemes (pixel-based, labeled as ``regressis,'' or object-based, labeled as ``catalog''), in the first redshift bin considered in the analysis ($0.8<z<2.1$). Right: Ratio of the weighted to unweighted power spectra. In this case, no significant deviation from $1$ is observed. The shaded grey area highlights the 1\% region around unity. }
    \label{fig:aic_kg}
\end{figure}

\subsubsection{Test on systematics-free mocks}
To evaluate the significance of mode removal using either the linear systematic correction weight or those creating using the linear regressor within the \texttt{regressis} package, we measure the angular power spectra of the mocks using each set of weights and compare them to the unweighted case (where all weights are set to unity). As the mocks do not have systematic contamination, a suppression of large-scale power in the weighted cases would indicate that the weights are spuriously removing cosmological signal instead of systematics, requiring a corresponding correction. We examine how the systematic weights impact both the galaxy auto-spectrum, $C_\ell^{gg}$, which is most sensitive to observational systematics and contributes to the covariance, and the cross-spectrum with CMB lensing, $C_\ell^{\kappa g}$, which is used in our main analysis. 
\begin{figure}[h!]
    \centering
    \includegraphics[width=0.75\linewidth]{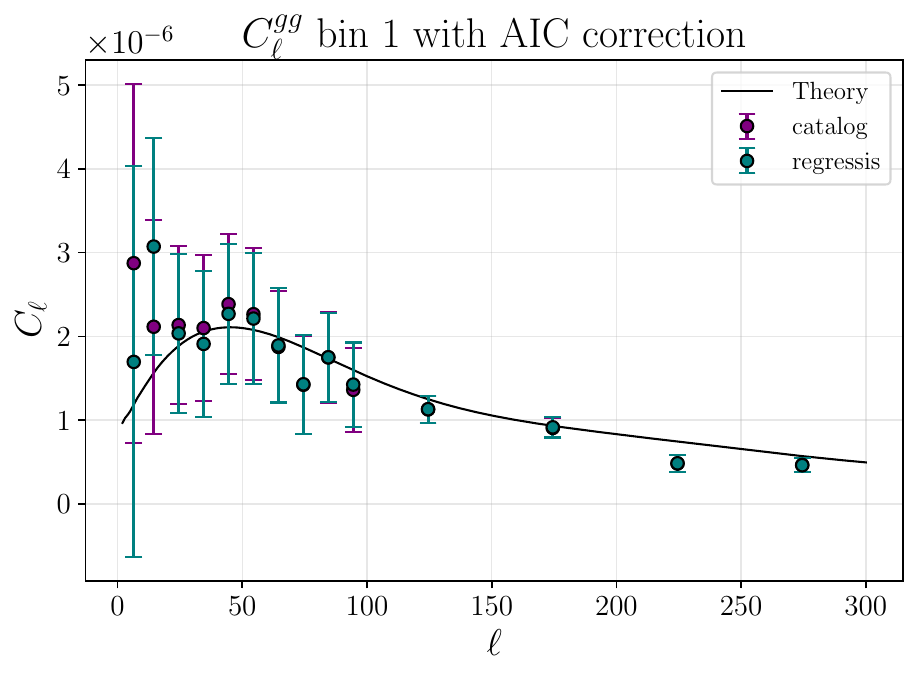}
    \caption{Auto-correlation angular power spectra measured from the catalog with the two possible sets of weights and corrected for mode removal. The error bars are computed from an analytical gaussian covariance computed using the \texttt{NaMaster} package, as discussed in Sec~\ref{sec:ps}. }
    \label{fig:clgg_bin1_corrected}
\end{figure}
\begin{figure}[h!]
    \centering
    \includegraphics[width=0.75\linewidth]{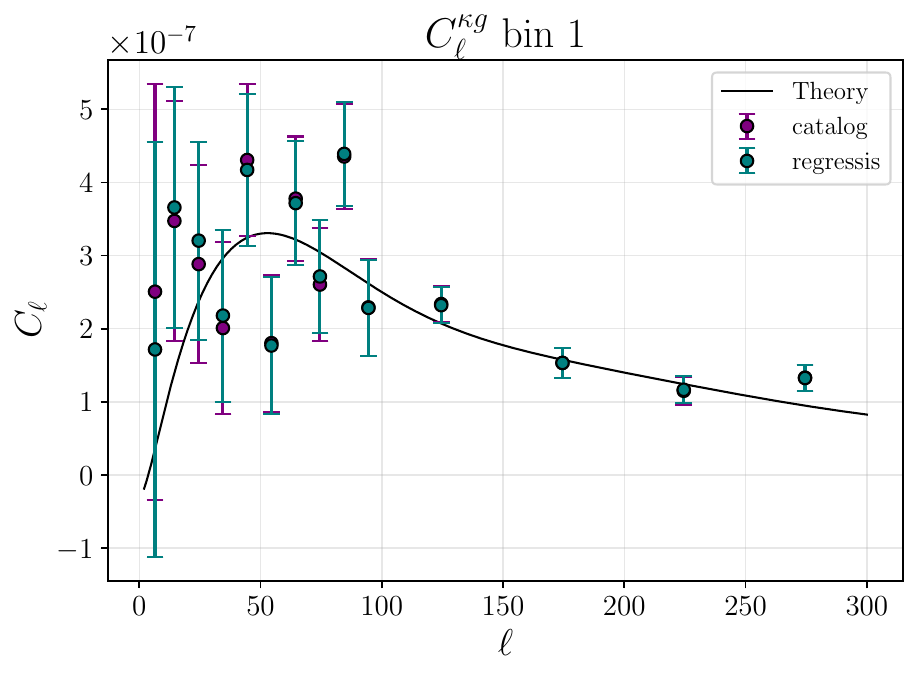}
    \caption{Cross-correlation angular power spectra measured from the catalog with the two possible sets of weights. In this case, no correction is applied as there is no sign of angular mode removal introduced in the cross correlation. The error bars come from the analytic estimation of the covariance matrix performed using the \texttt{NaMaster} package, described in Sec.~\ref{sec:ps}. }
    \label{fig:clkg_bin1_corrected}
\end{figure}
Fig.~\ref{fig:aic_gg} shows the results for the galaxy autocorrelation $C_\ell^{gg}$, indicating that both sets of weights are removing power on very large scales ($\ell<20$). We use the ratio, displayed on the right side of the plot, as a multiplicative correction to the power spectra measured on the data. The impact is more pronounced for the ``catalog'' linear weights, and the underlying cause is currently under investigation, as it must reside in the implementation of the weights.
Fig.~\ref{fig:clgg_bin1_corrected} shows the corrected autocorrelation measurement: we note that, even after accounting for angular mode removal, the auto power spectrum does not show signs of excess large scale power, a huge improvement with respect to the first analysis using the DESI DR9 Legacy Survey \citep{krolewski2024constraining}. This is can be attributed to the fact that the DESI DR1 spectroscopic catalog has a much higher purity compared to the photometric targets in the Legacy Survey and to the employment of the new catalog based estimator, which is more stable and robust to numerical instabilities and pixel related issues, as discussed in Sec.~\ref{sec:ps}. In this analysis we will adopt the more conservative and robust approach of using only the cross-correlation to constrain \fnl{}. Nonetheless, this result is encouraging as there is no need to model excess large scale power in the covariance, as in the previous analysis. Moreover, it indicates that we are going in the direction of being able to safely use the auto-correlation information for the \fnl{} analysis in future DESI data releases. 

Fig.~\ref{fig:aic_kg} shows no sign of over-subtraction of angular modes in the CMB lensing cross-correlation, which we employ in this work. Hence, we decided to not correct the measurement, an example of which is shown in Fig.~\ref{fig:clkg_bin1_corrected}.

\section{Optimal Redshift Weighting for $f_{\mathrm{NL}}$}
\label{sec:optw}
In the present work, we tested the use of optimal redshift weights for the angular two-point statistics, with the aim of improving constraining power of the analysis. The idea of optimal redshift weighting was first developed in \citep{Castorina18, Mueller19} and is now routinely applied in three-dimensional power spectrum analyses. For reference, the optimal \fnl{} weight for the galaxy monopole is given by \cite{Mueller17, Castorina18, Cagliari23}:
\begin{equation}
    w_{\hat{P}_0}(z) = w_{\mathrm{FKP}}^2\, b_\Phi(z)\, D(z)\, \left[b(z) + \frac{f(z)}{3}\right],
\end{equation}
where $w_{\mathrm{FKP}}$ is the standard FKP weight \citep{feldman1993power}. Since this is a weight for the power spectrum, the corresponding per-galaxy weight is its square root. 
\begin{figure}[h!]
    \centering
    \includegraphics[width=0.95\linewidth]{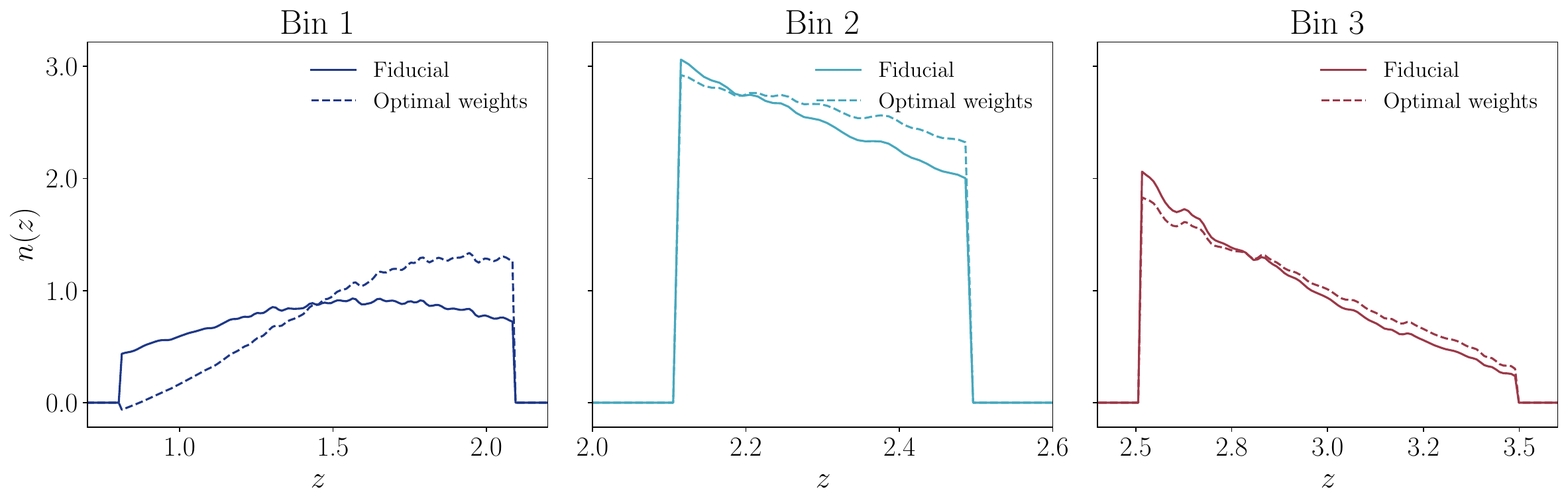}
    \caption{Redshift distribution of the DESI DR1 quasars. The plot shows the $n(z)$ of the three tomographic bins in which the data is divided for this analysis, in order of increasing redshift. The solid lines represent the fiducial $n(z)$, and the dashed lines show the redshift distribution after weighting the galaxies with the weights in Eq.~\ref{eqn:optw}. The displayed redshift distributions are normalized in each bin.}
    \label{fig:nz_weights}
\end{figure}

In practice, this corresponds to cross-correlating two weighted fields, respectively proportional to $w_{\mathrm{FKP}}(b(z) - p)$ and $w_{\mathrm{FKP}}\!\left[b(z) + \frac{f(z)}{3}\right]$, as in \cite{Cagliari23}. This weighting scheme enhances the constraining power on \fnl{} by explicitly accounting for the redshift dependence of the PNG signal.
We extend this framework to the angular cross-correlation between CMB lensing and galaxy overdensity, $C_\ell^{\kappa g}$, and derive an optimal estimator for this observable. The details of the derivation  are given in App.~\ref{app:opt_weights_derivation}. We find that the optimal redshift weight that maximizes the \fnl{} sensitivity of $C_\ell^{\kappa g}$ is:
%Briefly, the first term of Eq.~\ref{eqn:optw_def} yields \(\frac{\partial C_\ell^{\kappa g}}{\partial f_\mathrm{NL}} = \frac{W_\kappa(\chi)}{\chi^2}\,(b(z) - p)\,D(z),\) while the inverse covariance term contributes the expected dependence on the FKP weights, as all the redshift independent terms drop when normalizing the weights as in \citep{Mueller17}. Combining these results, the optimal redshift weight that maximizes the \fnl{} sensitivity of $C_\ell^{\kappa g}$ is:
\begin{equation}\label{eqn:optw}
    w_\mathrm{opt}(z) = (b(z)-p)\,w_{\mathrm{FKP}},
\end{equation}
normalized such that $\int w_\mathrm{opt}(z)\,\mathrm{d}z = 1$. As in the power spectrum case, the resulting weights effectively up-weight higher-redshift galaxies, where the \fnl{} signature is strongest. As one might expect, we only get one power of the FKP weight here, as opposed to the two powers in the power spectrum case. 
\begin{figure}[h!]
    \centering
    \includegraphics[width=0.7\linewidth]{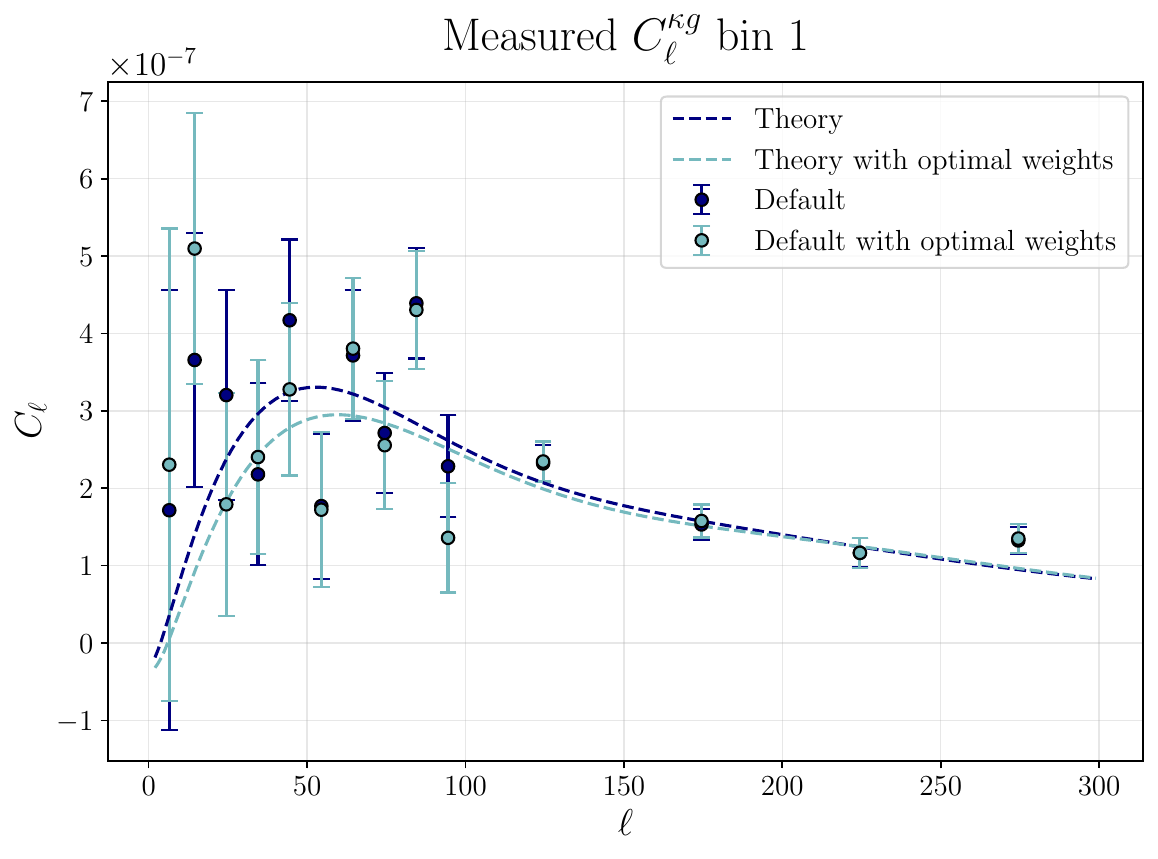}
    \caption{$C_\ell^{\kappa g}$ in the first tomographic bin. In navy, the measurement and theory curve in the default configuration (i.e., using the \texttt{regressis} linear weights, see Sec.~\ref{sec:sys_weights}). In light green, the case where the optimal weights are applied. This plot showcases the impact of the optimal weights on the pipeline.}
    \label{fig:cl_weights}
\end{figure}

The implementation of these weights is straightforward. We evaluate the optimal weights for each galaxy in the catalog using the expressions derived above. The CMB lensing kernel $W_\kappa$ is defined in Eq.~\ref{eqn:cmb_kernel}, while $b_\Phi$ is given in Eq.~\ref{eqn:bphi_sims}, with the linear bias $b(z)$ following the model of \citep{chaussidon2024}. The growth factor $D(z)$ is computed using the Boltzmann solver \texttt{CAMB} \citep{lewis2000efficient}, and the FKP weights \citep{feldman1993power} are already provided as an attribute of the galaxy catalog. In the default computation of the redshift distribution $n(z)$, each galaxy is weighted according to Eq.~\ref{eqn:galaxy_weights} before binning. When including the optimal weighting scheme, the total weight assigned to each galaxy becomes
\(
    w = w_\mathrm{tot}\, w_\mathrm{opt} = w_\mathrm{comp}\, w_\mathrm{sys}\, w_\mathrm{zfail}\, w_\mathrm{opt},
\)
effectively modifying the inferred $n(z)$. As shown in Fig.~\ref{fig:nz_weights}, the optimal weights up-weight the high-redshift part of the sample, where the sensitivity to \fnl{} is greatest. This occurs because galaxies at higher redshift are more affected by a potential PNG signal, which enhances their large scale clustering. On the other hand, at the lowest redshifts the weights become negative. In the first redshift bin (in particular in the range $0.8 < z < 0.9$), some galaxies receive negative weights. This is expected and does not pose any issue for the analysis: the negative values simply indicate that galaxies in this range would respond to a stronger PNG signal with a reduction in clustering amplitude. Measuring \fnl{} from such an absence of clustering is intrinsically more difficult, and the optimal weighting appropriately accounts for this effect. 

To apply the optimal weights consistently throughout the analysis, we must consider all components of the pipeline that are affected. A change in the galaxy redshift distribution modifies the theoretical predictions, since $n(z)$ enters the number counts kernel $W_\delta(\chi)$ defined in Eq.~\ref{eqn:number_counts}. Consequently, the theory curves used in the analytic covariance computation (see Sec.~\ref{sec:ps}) must also be updated. The analysis performed with optimal weights therefore employs a covariance matrix recomputed following the same procedure described in Sec.~\ref{sec:ps}, but using the weighted $n(z)$. Finally, the weighting scheme must also be applied to the data when measuring the angular power spectra. Fig.~\ref{fig:cl_weights} illustrates the effect of the optimal weights on the measured $C_\ell^{\kappa g}$ and on the corresponding theoretical predictions. The comparison between the default configuration (defined in Sec.~\ref{sec:sys_weights}) and the weighted case highlights how both the measurement and the theoretical model respond to the modified redshift distribution.

\section{Results and Discussion}
\label{sec:results}
\begin{figure}[h!]
    \centering
    \includegraphics[width=\linewidth]{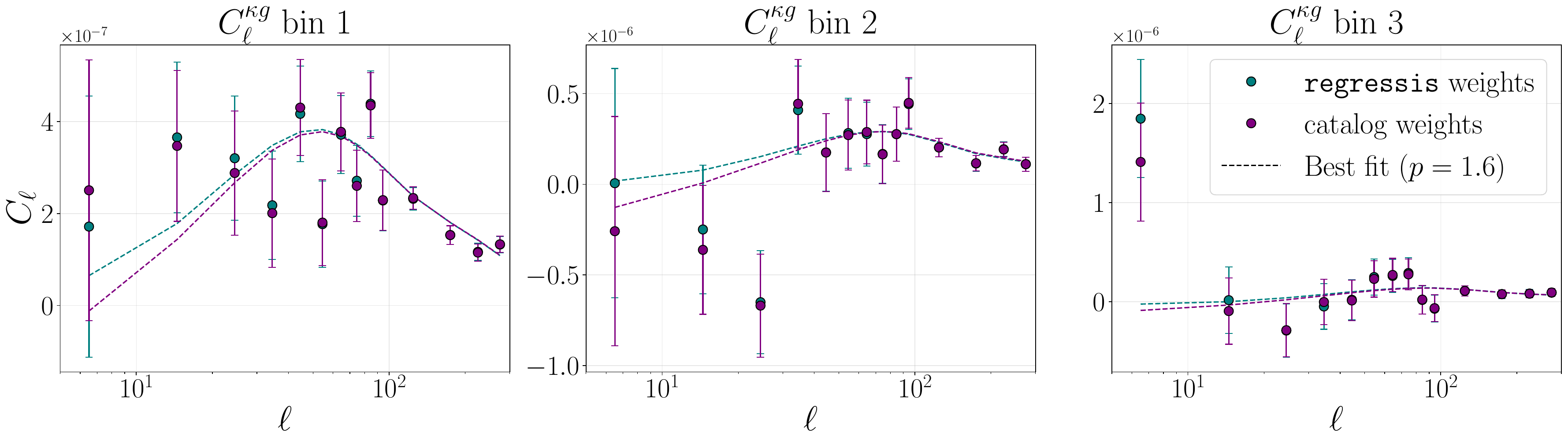}
    \caption{As in Fig.~\ref{fig:clkg_bin1_corrected}, the scatter points represent the measurements on the data performed with two possible weighting schemes: the default choice \texttt{regressis} in green, and the alternative linear weighting scheme implemented in the DESI catalog in purple. The dashed lines represent the corresponding best fit curves when assuming $p=1.6$. This baseline result does not utilize the optimal weighting scheme.}
    \label{fig:bestfits}
\end{figure}

In this section we present tomographic constraints on primordial non-Gaussianity obtained from the cross-correlation between the \Planck PR4 CMB lensing convergence maps and spectroscopically confirmed quasars from DESI DR1. Our analysis covers the redshift range $0.8 \le z \le 3.5$, divided into three tomographic bins, and includes a total of $1{,}223{,}391$ quasars across $7{,}200~\mathrm{deg}^2$ of sky. We use the angular power spectrum estimator described in Sec.~\ref{sec:ps} to measure the cross-spectra $C_\ell^{\kappa g}$. Our parameter inference is performed with \texttt{Turing.jl}\footnote{\url{https://github.com/TuringLang/Turing.jl}} \citep{turing1}, which provides an interface to define probabilistic models in terms of explicit priors and our differentiable \texttt{Blast.jl} likelihood. Posterior exploration is carried out using the No-U-Turn Sampler (NUTS) algorithm \citep{hoffman2014no}, a self-tuning Hamiltonian Monte Carlo sampler that efficiently leverages gradient information and has been widely adopted for cosmological inference. In this baseline configuration, we do not employ the optimal weighting scheme, which we instead treat as an analysis variation in Sec.~\ref{sec:optw_tests}.
The main results is presented in Table~\ref{tab:main_results}, visualized in Fig.~\ref{fig:bestfits} and discussed in the following text. Throughout this section we also present results for various analysis variations and tests that we performed, like the inclusion of optimal \fnl{} weights and different redshift cuts and bins. Everything is summarized in Fig.~\ref{fig:comparison}, showcasing the robustness of the analysis.
As discussed in Sec.~\ref{sec:intro}, uncertainties in the theoretical prediction of the non-Gaussian bias coefficient $b_\Phi$ create a degeneracy between $f_{\mathrm{NL}}$ and $b_\Phi$. In general, one could constrain the product $f_{\mathrm{NL}}\,b_\Phi$ directly, without assuming a specific relation between $b_\Phi$ and $b_1$ \citep[see e.g.][]{barreira2022can}, since a detection of the product would remain physically meaningful. Our goal here, however, is not to explore this broader parameterization, but to assess how the inferred $f_{\mathrm{NL}}$ shifts under reasonable choices for the $b_\Phi(b_1)$ relation. We therefore consider two benchmark cases commonly used in the literature \citep[see e.g.][]{krolewski2024constraining, chaussidon2024, fabbian2025}: $p=1.6$, corresponding to a recent-merger scenario, and $p=1.0$, corresponding to the universality assumption. We impose flat priors on all parameters, with $f_{\mathrm{NL}} \sim \mathcal{U}(-200,200)$ and $b_i \sim \mathcal{U}(0,2)$.

To assess the robustness of our results, we performed a series of analysis variations and validation tests, described below. Fig.~\ref{fig:bestfits} presents the measured angular power spectra together with the best-fit model for the two sets of systematic weights considered. In the third tomographic bin, the lowest multipole ($\ell=6.5$) shows a deviation of $\approx 2\sigma$ from the best-fit prediction. This feature is consistent with statistical noise and does not dominate the total $\chi^2$. However, we know that the third bin is the one where the quasar sample has lower purity, so it is possible that some remaining systematic in the CMB lensing maps could correlate with uncorrected systematics in the sample resulting in the large scale extra power. For this reason, we run the analysis in the default configuration removing that point, which is located at scales very relevant to constrain \fnl{}. The result in this case is $f_\mathrm{NL}=-22^{+27}_{-34}$ as opposed to $f_\mathrm{NL}=2^{+28}_{-34}$ when all the points are considered. The error bars stay consistent, but there is a $\approx 0.8\sigma$ shift in the best fit value. We decided to keep the analysis as it is and not exclude that point from the pipeline as the evidence for doing that is not compelling enough to avoid falling into confirmation bias. 

\begin{figure}[h!]
    \centering
    \begin{subfigure}{0.49\textwidth}
        \centering
        \includegraphics[width=\linewidth]{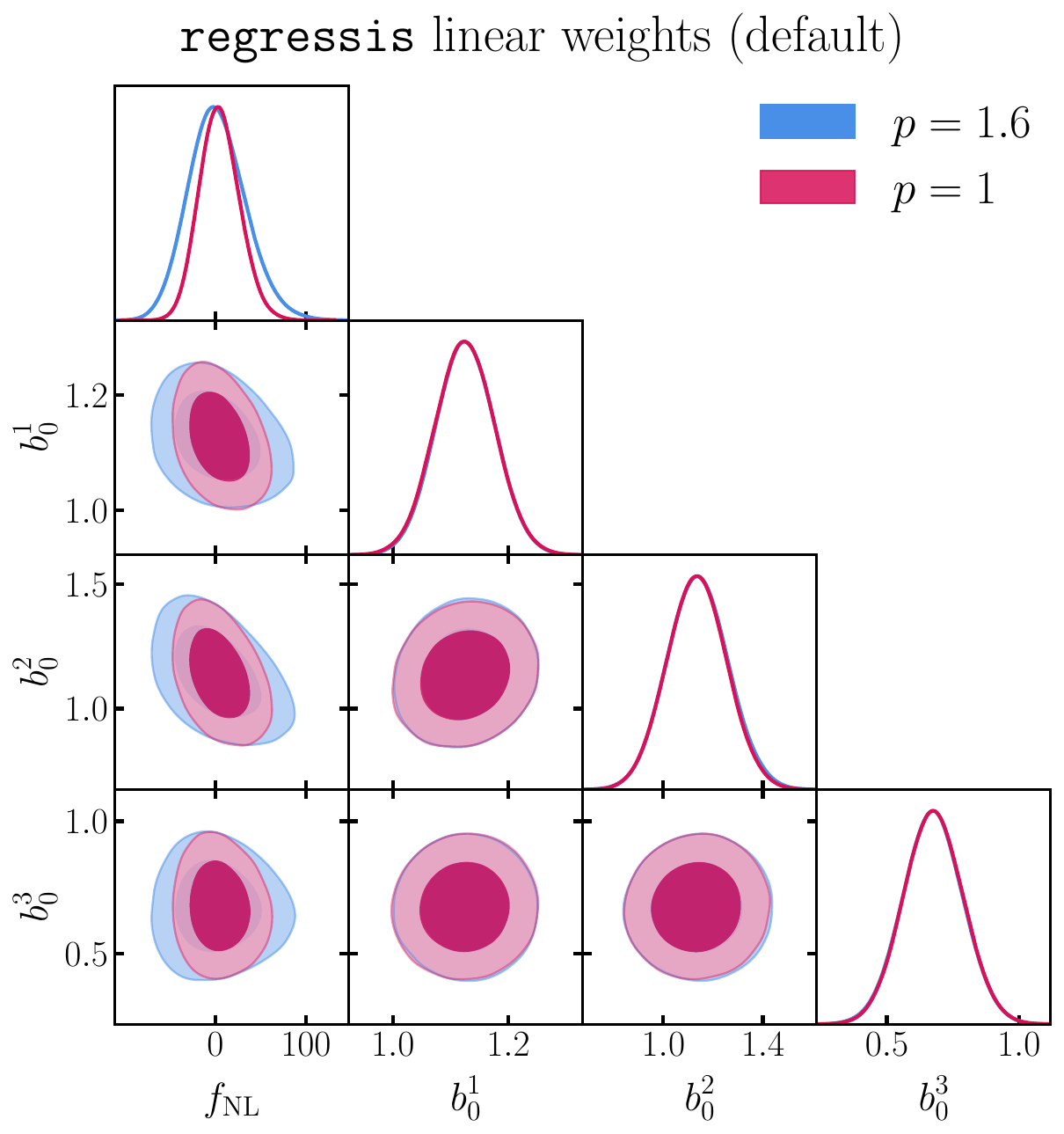}
    \end{subfigure}
    \begin{subfigure}{0.49\textwidth}
        \centering
        \includegraphics[width=\linewidth]{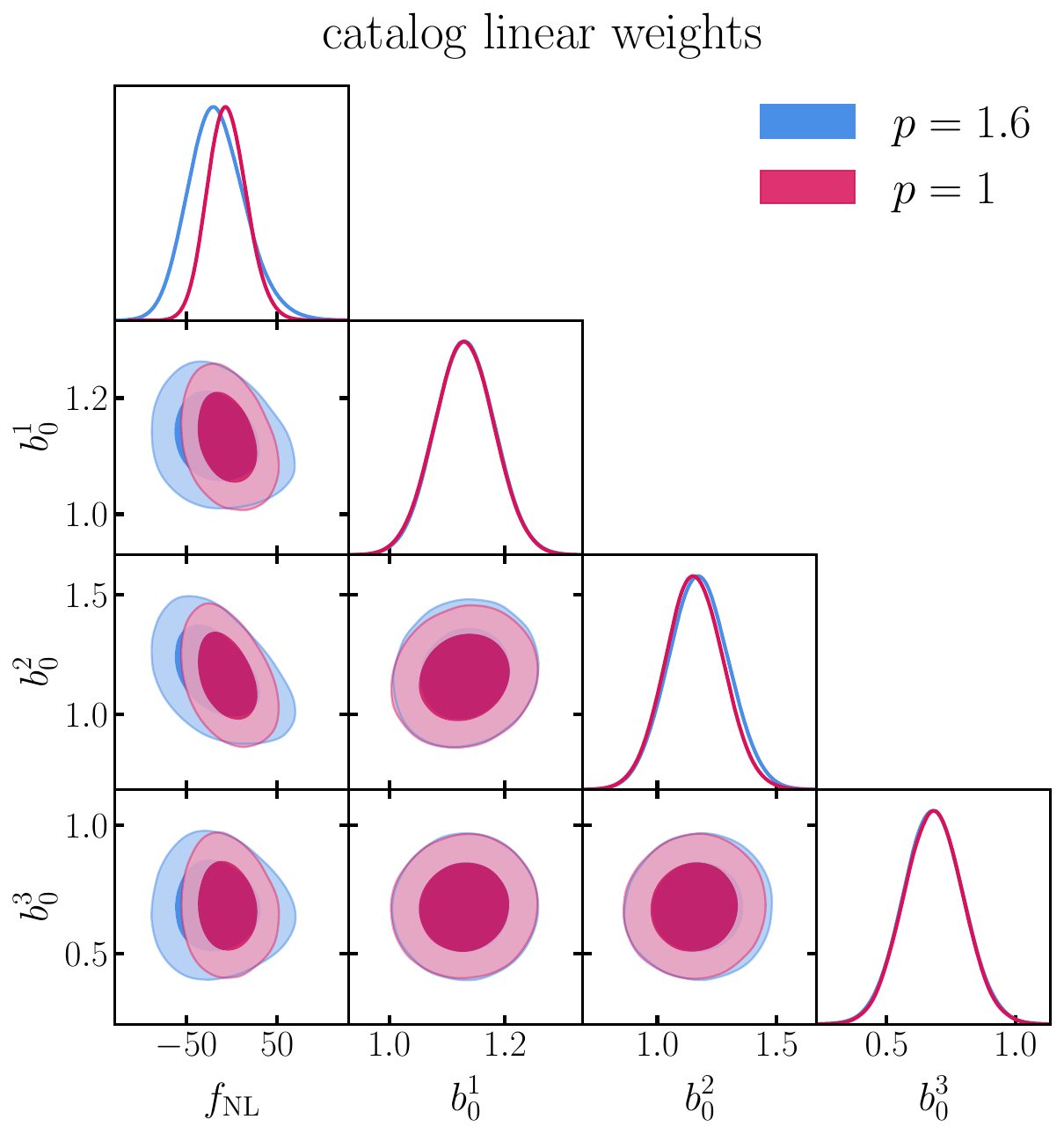}
    \end{subfigure}
    \caption{Corner plots showing the constraints for the $4$ free parameters in the analysis, namely \fnl{} and a bias amplitude for each of the $3$ tomographic bins. \emph{Left}: constraints in the fiducial configuration, using the \texttt{regressis} linear weights. \emph{Right}: results of the analysis performed with a different choice of systematic weights, \textit{i.e.} the ones implemented in the official DESI catalog.}
    \label{fig:contours_fid}
\end{figure}

The marginalized constraints corresponding to each configuration are summarized in Table~\ref{tab:main_results}. The best-fit values are consistent within $1\sigma$ across all weighting schemes, and the reduced $\chi^2$ values confirm the overall quality of the fits. If we use $p=1$ instead of $p=1.6$, the constraints become tighter: this is to be expected as, in that case, the quasars are more sensitive to variation in \fnl{}. The same results can be visualized in Fig.~\ref{fig:contours_fid}, which shows the contours for the $4$ free parameters. The constraints on the bias scaling amplitudes $b_0^i$ are not affected by variations in the analysis: Table~\ref{tab:biases_3b} shows the marginalized posteriors for the bias amplitudes for the two possible choices of systematic weights. Those results have been obtained with $p=1.6$. Interestingly, the bias amplitude that we fit for the third bin comes out to be $\approx 20-30\%$ lower than expected, with a best fit value of $b_0^3=0.68$, very robust to variations in the analysis. This discrepancy has emerged in other DESI analyses \citep{de2025cosmology}, however the reason is still unknown and is currently being investigated. Nonetheless, since this fact can raise doubts on the validity of the assumed bias relation \citep{chaussidon2024}, we test the impact of that choice in Sec.~\ref{sec:bias_test}.

\begin{figure}[h!]
    \centering
    \includegraphics[width=0.7\linewidth]{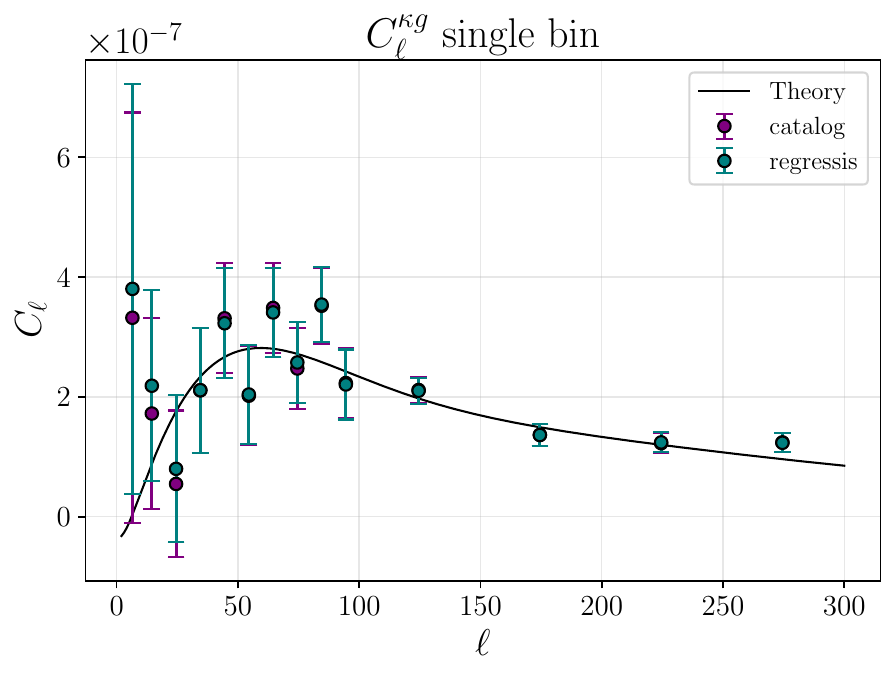}
    \caption{$C_\ell^{\kappa g}$ when the quasars are not divided into tomographic bins.}
    \label{fig:sb_clkg}
\end{figure}

\begin{table}[h!]
    \centering
    \renewcommand{\arraystretch}{1.3} % increases row height
    \setlength{\tabcolsep}{12pt} % increases column spacing
    \begin{tabular}{|c|c|c|}
        \hline
        $p$ & \texttt{regressis} linear weights & catalog linear weights\\
        \hline\hline
        $1.6$ & $f_{\mathrm{NL}} = 2^{+28}_{-34}\,\, (\chi^2=56)$  & $f_{\mathrm{NL}} = -14^{+27}_{-33}\,\, (\chi^2=53)$ \\
        \hline
        $1.0$ & $f_{\mathrm{NL}} = 6^{+20}_{-24}\,\, (\chi^2=54)$  & $f_{\mathrm{NL}} = -4^{+19}_{-22}\,\, (\chi^2=53)$ \\
        \hline
    \end{tabular}
    \caption{Constraints on $f_{\mathrm{NL}}$ for different choices of $p$ and weighting schemes (calculated without the additional optimal \fnl{} weights described in Sec.~\ref{sec:optw}. The results from the fiducial scenario are reported in the top left corner. The effective number of degrees of freedom in the analysis is $38$, giving reduced $\chi^2$ values of $1.47$ and $1.39$ respectively. }
    \label{tab:main_results}
\end{table}
\begin{table}[h!]
    \centering
    \renewcommand{\arraystretch}{1.3} % increases row height
    \setlength{\tabcolsep}{12pt} % increases column spacing
    \begin{tabular}{|c|c|c|}
    \hline
         &  \texttt{regressis} linear weights& catalog linear weights\\
         \hline\hline
         $b_0^1$&  $1.13 \pm 0.05$& $1.13 \pm 0.05$ \\
         \hline
         $b_0^2$&  $1.14 \pm 0.13$& $1.17 \pm 0.12$\\
         \hline
         $b_0^3$&  $0.68 \pm 0.12$& $0.68 \pm 0.12$\\
         \hline
    \end{tabular}
    \caption{Bias amplitude constraints for the two possible choices of linear weights tested throughout this paper. The default choice are the \texttt{regressis} linear weights, but the weight choice does not have an impact on the bias amplitudes as shown in this table.}
    \label{tab:biases_3b}
\end{table}
To test the impact of the tomographic redshift bins on the constraints, we also performed the analysis using a single redshift bin $0.8<z<3.5$. The measurement on the data in this case is displayed in Fig.~\ref{fig:sb_clkg}, while the marginalized constraints for \fnl{} are reported in Table~\ref{tab:singlebin}. The constraining power in this case deteriorates, indicating how the redshift binning or the inclusion of optimal weights, as discussed in Sec.~\ref{sec:optw_tests}, is important in extracting \fnl{} information. 
\begin{table}[h!]
    \centering
    \renewcommand{\arraystretch}{1.3} % increases row height
    \setlength{\tabcolsep}{12pt} % increases column spacing
    \begin{tabular}{|c|c|c|}
        \hline
        $p$ & \texttt{regressis} linear weights & catalog linear weights\\
        \hline\hline
        $1.6$ & $f_{\mathrm{NL}} = 17^{+30}_{-40}\,\, (\chi^2=11)$  & $f_{\mathrm{NL}} = 4^{+33}_{-41}\,\, (\chi^2=11)$ \\
        \hline
        $1.0$ & $f_{\mathrm{NL}} = 12^{+24}_{-28}\,\, (\chi^2=11)$  & $f_{\mathrm{NL}} = 3^{+23}_{-27}\,\, (\chi^2=11)$ \\
        \hline
    \end{tabular}
    \caption{Constraints on $f_{\mathrm{NL}}$ for the analysis without tomographic bins. Again, different choices of $p$ and weighting schemes have been tested and the results from the fiducial scenario are reported in the top left corner. The effective number of degrees of freedom in the analysis is $12$, giving reduced $\chi^2$ values of $0.91$. This corresponds to a $p$-value of $0.5$, indicating that the model provides a good fit to the data. }
    \label{tab:singlebin}
\end{table}

\subsection{The optimal weights results}\label{sec:optw_tests}
In existing literature \citep{Cagliari23, Castorina, chaussidon2024, Mueller} about optimal weighting schemes to maximize the \fnl{} signal, it was found that the improvement of constraining power coming from this technique is around $8-10\%$ in the case of the 3D power spectrum. To test this in our own pipeline, we compare our baseline results to the case where the optimal weighting scheme is also applied. As the marginalized constraints in Table~\ref{tab:optw_results} show, we find a similar gain ($\approx 5-10\%$), however we also observe relevant shifts in the best fit values. We also test the impact of optimal weights in the single bin analysis, whose results are displayed in Table~\ref{tab:optw_sb_results}: while the error bars only marginally improve, again the best fit values for \fnl{} suffer from quite significant shifts (from $0$ to $20$, $\approx 0.7\sigma$, and from $-16$ to $10$, $\approx 0.8\sigma$). Despite the substantial shifts in the posterior means, the mean obtained with optimal weighting falls within the $1\sigma$ range of our default analysis. Nonetheless, these shifts suggest that the data is quite noisy (as illustrated in Fig.~\ref{fig:bestfits}) and characterized by large error bars. This noise could potentially obscure the advantages of employing optimal weights. Furthermore, given that the \fnl{} signal is dependent on redshift and that we are dividing our dataset into three tomographic bins, it is possible we are already extracting the majority of the available information.  
\begin{table}[h!]
    \centering
    \renewcommand{\arraystretch}{1.3} % increases row height
    \setlength{\tabcolsep}{12pt} % increases column spacing
    \begin{tabular}{|c|c|c|}
        \hline
        Optimal weights & \texttt{regressis} linear weights & catalog linear weights\\
        \hline\hline
        No & $f_{\mathrm{NL}} = 2^{+28}_{-34}\,\, (\chi^2=56)$  & $f_{\mathrm{NL}} = -14^{+27}_{-33}\,\, (\chi^2=53)$ \\
        \hline
        Yes & $f_{\mathrm{NL}} = 19^{+25}_{-31}\,\, (\chi^2=62)$  & $f_{\mathrm{NL}} = 9^{+25}_{-31}\,\, (\chi^2=59)$ \\
        \hline
    \end{tabular}
    \caption{Comparison of the results with and without applying the optimal weighting scheme. The fiducial value of $p=1.6$ is assumed. }
    \label{tab:optw_results}
\end{table}
\begin{table}[h!]
    \centering
    \renewcommand{\arraystretch}{1.3} % increases row height
    \setlength{\tabcolsep}{12pt} % increases column spacing
    \begin{tabular}{|c|c|c|}
        \hline
        Optimal weights & \texttt{regressis} linear weights & catalog linear weights\\
        \hline\hline
        No & $f_{\mathrm{NL}} = 17^{+30}_{-40}\,\, (\chi^2=14)$  & $f_{\mathrm{NL}} = 4^{+33}_{-41}\,\, (\chi^2=14)$ \\
        \hline
        Yes & $f_{\mathrm{NL}} = 10^{+28}_{-35}\,\, (\chi^2=14)$  & $f_{\mathrm{NL}} = -4^{+28}_{-36}\,\, (\chi^2=14)$ \\
        \hline
    \end{tabular}
    \caption{Comparison of the results with and without applying the optimal weighting scheme for the analysis with a single tomographic bin. The fiducial value of $p=1.6$ is assumed. }
    \label{tab:optw_sb_results}
\end{table}

\begin{figure}[h!]
    \centering
    \begin{subfigure}{0.49\textwidth}
        \centering
        \includegraphics[width=\linewidth]{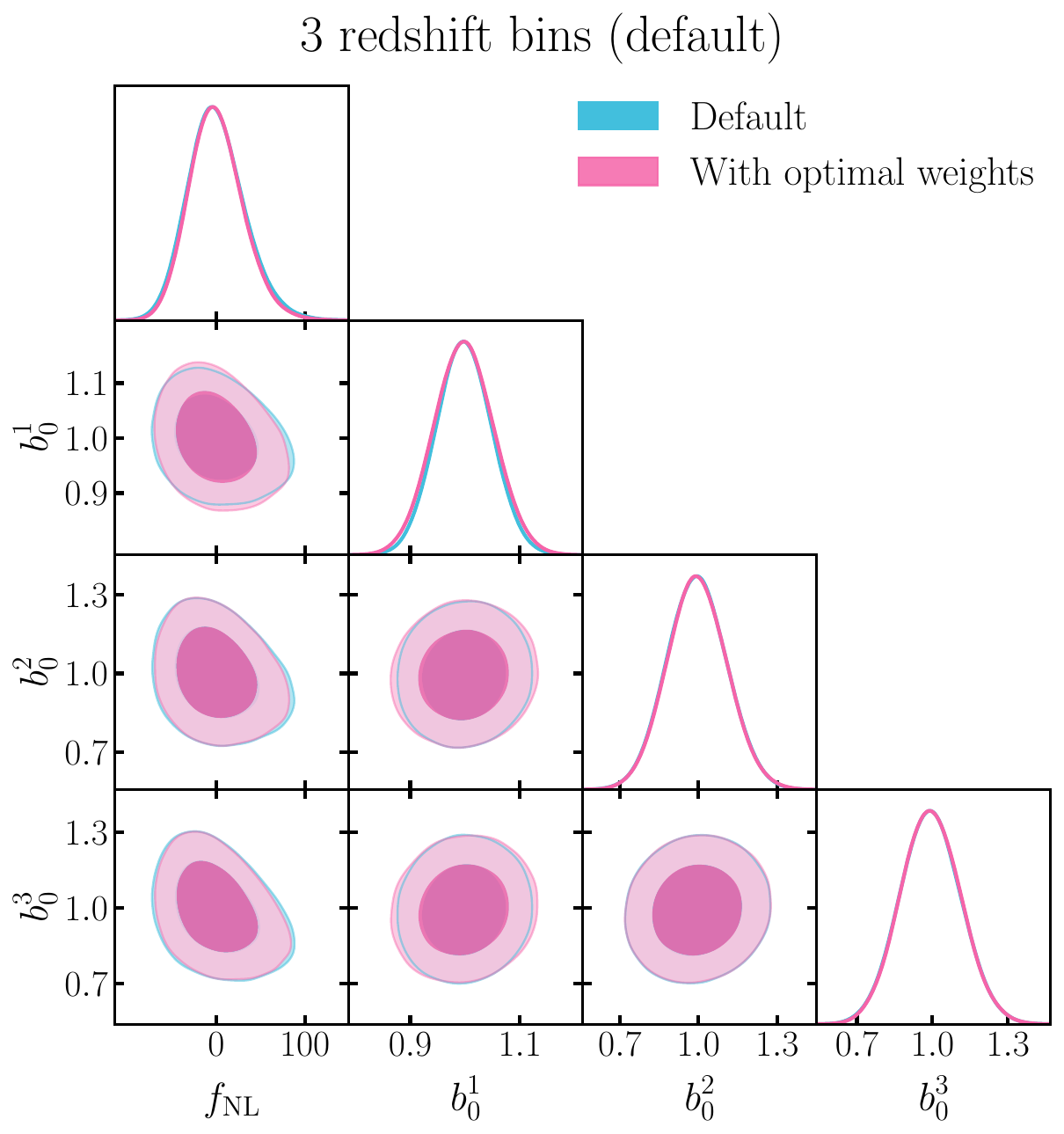}
    \end{subfigure}
    \hspace{0.005\textwidth}
    \begin{subfigure}{0.49\textwidth}
        \centering
        \includegraphics[width=\linewidth]{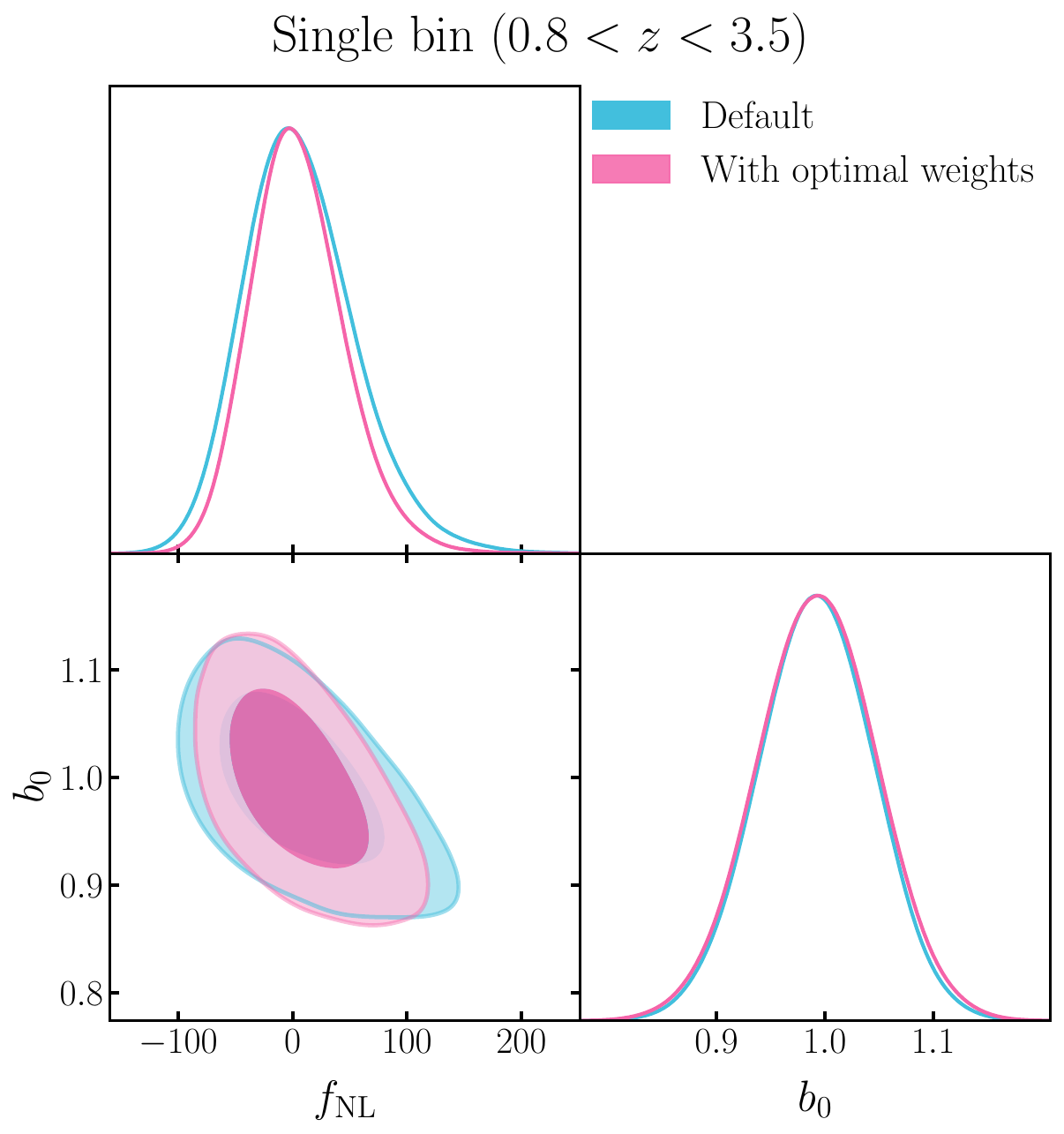}
    \end{subfigure}
    \caption{Results of the test using a noiseless data vector to assess the performance of the optimal weights. Left: Analysis in the default configuration, where the dataset is split into three redshift bins. In this case, the optimal weights (pink contours) do not improve the \fnl{} constraint relative to the baseline (blue). Right: Results of the analysis performed without tomography, grouping the full quasar sample into a single bin. Here, the error bars on \fnl{} are $\approx 10\%$ tighter when applying the optimal weights (pink) compared to the unweighted case (blue). In both panels, \texttt{regressis} linear weights with $p = 1.6$ were used.}
    \label{fig:contours_optw_test}
\end{figure}

To investigate these hypotheses regarding the performance of the optimal weights, we conducted the analysis using a synthetic noiseless data vector created with $f_\mathrm{NL}=0$ and $b_0^i=1$. This was done in two different setups: the default case with $3$ tomographic redshift bins, and a simplified case without tomography, which utilized a single redshift bin ranging from $z=0.8$ to $z=3.5$. We tested the impact of the optimal weights in these two scenarios, obtaining the constraints reported in Table~\ref{tab:optw_test} and visualized in Fig.~\ref{fig:contours_optw_test}. The results for the bias amplitudes are not reported because, as expected, the optimal weights for \fnl{} do not affect the bias constraints, which are completely consistent to that reported in Table~\ref{tab:biases_3b} for the analysis without optimal weights.
\begin{table}[h!]
    \centering
    \renewcommand{\arraystretch}{1.3} % increases row height
    \setlength{\tabcolsep}{12pt} % increases column spacing
    \begin{tabular}{|c|c|c|}
        \hline
        Optimal weights & $3$ redshift bins (default) & Single bin $(0.8<z<3.5)$ \\
        \hline\hline
        No & $f_{\mathrm{NL}} = 0^{+28}_{-35}$  & $f_{\mathrm{NL}} = 0^{+40}_{-54}$ \\
        \hline
        Yes & $f_{\mathrm{NL}} = 0^{+25}_{-32}$  & $f_{\mathrm{NL}} = 0^{+35}_{-42}$ \\
        \hline
    \end{tabular}
    \caption{Marginalized constraints for \fnl{} when assessing the impact of the tomographic bins and noisy data on the optimal weights effectiveness.  }
    \label{tab:optw_test}
\end{table}

The results of this test are very interesting: in fact, in the case of the tomographic analysis, when using noiseless data the optimal weights are providing a $ \approx 5-10\%$ improvement in the constraining power, similar to what we see in the analysis with real data. On the other hand, when we employ only one redshift bin, then using optimal weights tightens the constraints by $\approx 15-20\%$ in this noiseless data vector test. This suggests that, by binning our data, we are already extracting most of the redshift information that helps in better constraining \fnl{}. Overall, the optimal weighting scheme is straightforward to implement and computationally inexpensive, making it a promising extension of the analysis. However, since the tomographic approach already achieves comparable performance and the current measurements are still limited by systematic uncertainties, the practical gain from applying optimal weights is modest. For this reason, we adopt the results without optimal weighting as our fiducial choice. Looking ahead, the method is expected to become increasingly valuable for future DESI data releases and other Stage-IV surveys, where a $\sim10-15\%$ improvement in constraining power could play a significant role in reaching the target precision of $\sigma_{f_\mathrm{NL}} < 1$.

\subsection{Fitting a single bias amplitude for all three bins}
As an alternative to fitting a bias amplitude per tomographic bin, we tried to fit the same amplitude $b_0$ to all three bins. The resulting marginalized \fnl{} constraints are presented in the second row of Table~\ref{tab:1biasfit}. These constraints are compared with the fiducial results. 
\begin{table}[h!]
    \centering
    \renewcommand{\arraystretch}{1.3} % increases row height
    \setlength{\tabcolsep}{12pt} % increases column spacing
    \begin{tabular}{|c|c|c|}
        \hline
        Bias & \texttt{regressis} linear weights & catalog linear weights\\
        \hline\hline
        3 biases & $f_{\mathrm{NL}} = 2^{+28}_{-34}\,\, (\chi^2=56)$  & $f_{\mathrm{NL}} = -14^{+27}_{-33}\,\, (\chi^2=53)$ \\
        \hline
        1 bias & $f_{\mathrm{NL}} = -3^{+25}_{-28}\,\, (\chi^2=70)$  & $f_{\mathrm{NL}} = -20\pm 25\,\, (\chi^2=66)$ \\
        \hline
    \end{tabular}
    \caption{Results for the test of fitting a single bias amplitude $b_0$ for all three bins, compared to the baseline configuration where we allow for each redshift bin to have a free amplitude $b_0^i$. When fitting a single bias amplitude the reduced $\chi^2$ rises from $1.47$ to $1.70$, indicative of a worst fit. The results reported here were obtained using $p=1.6$ and without employing the optimal weights.}
    \label{tab:1biasfit}
\end{table}
The resulting error bars on \fnl{} are tighter because the analysis has more degrees of freedom ($40$ instead of $38$); however, as indicated by the increase in the reduced $\chi^2$ from $1.4$ to $1.7$, indicating that this fit is not as good as the fiducial one. This issue arises because the third tomographic bin has been observed to prefer a much lower amplitude ($b_0^3=0.68$ compared to $b_0^{1,2}=1.1$), making it difficult to accurately represent this with a single amplitude fit. In fact the best fit values for $b_0$ in this scenario are $b_0=1.07\pm0.05$ and $b_0=1.08\pm 0.05$ for the \texttt{regressis} and catalog weights respectively. The decision of having a free amplitude per tomographic bin as our fiducial configuration was driven by this goodness-of-fit test, and for consistency with other similar analyses \citep{de2025cosmology, fabbian2025}. Nevertheless, the notably low amplitude in the third bin demonstrates that our baseline bias model does not fully capture the clustering properties of the quasars, thereby motivating the additional tests presented in Sec.~\ref{sec:bias_test}. 

\subsection{Impact of the bias relation}\label{sec:bias_test}
The baseline analysis reports some inconsistencies emerging with the third tomographic bin: first of all, the lowest multipole $\ell=6.5$ shows a $>2\sigma$ deviation from the best fit model which might be a sign of some remaining uncorrected systematic. Moreover, the marginalized posterior for the bias amplitude $b_0^3$ is not compatible with $1$ at $\approx 3\sigma$ level, indicating that the bias we assumed is incorrect for high redshift quasars. Despite this discrepancy emerging in other DESI analysis, it is important to test its impact on the \fnl{} marginalized posterior, the focus of this work. 
As a first test, we tried to completely remove the third bin from the analysis, performing it with only the first two tomographic bins ($0.8<z<2.1$, and $2.1<z<2.5$). The resulting marginalized constraint is $f_\mathrm{NL}=-15^{+29}_{-35}$. Noticeably, the best fit value shifts by $\approx 0.5\sigma$ compared to the baseline result ($f_\mathrm{NL}=2^{+28}_{-34}$), but the constraining is almost unchanged, suggesting that the third bin could be removed from the analysis without any substantial loss of information. As this was discovered later in the analysis, we do not change our baseline to this case to avoid confirmation bias and to stay consistent with the analysis presented in \citep{de2025cosmology}.

The marginalized constraints for the bias amplitudes $b_0^i$, reported in Table~\ref{tab:biases_3b}, are indicative of the fact that the bias evolution we choose does not represent well the bias of our QSO sample. In order to better describe our sample, we tested two alternative bias relations: we find the best fit effective bias of each tomographic bin by evaluating our bias model at the effective redshift of each bin (those values are reported in Table~\ref{tab:DESI_DR1_data}), with $b_0^i$ being the marginalized posteriors (reported in Table~\ref{tab:biases_3b}). The resulting values are: $b(z_\mathrm{eff}^1)=2.46$, $b(z_\mathrm{eff}^2)=3.77$, $b(z_\mathrm{eff}^3)=2.79$. Starting from these values, we construct a constant bias model: in each bin, the bias is constant and set to the values just reported. Alternatively, we define a redshift dependent bias model $b(z)$ by linearly interpolating between the three values. 
\begin{table}[h!]
    \centering
    \renewcommand{\arraystretch}{1.3} 
    \setlength{\tabcolsep}{12pt}
    \begin{tabular}{|c|c|c|}
    \hline
       Parameter & Constant $b(z)$ &  Linear Interpolation $b(z)$ \\
       \hline
       $f_\mathrm{NL}$ & $-3^{+28}_{-35}$ & $2^{+28}_{-33}$\\
       $b_0^1$ & $1.01\pm0.05$ & $1.02\pm0.05$\\
       $b_0^2$ & $1.12\pm0.12$& $1.16\pm0.13$\\
       $b_0^3$ & $1.07\pm0.17$& $1.16\pm0.18$\\
         \hline
    \end{tabular}
    \caption{Constraints on the parameters obtained from the analyses performed with different prescriptions for the bias model $b(z)$. In both cases, the reduced $\chi^2$ of the fit is $1.5$}
    \label{tab:diff_bias}
\end{table}
\begin{figure}[h!]
\centering
    \includegraphics[width=0.7\linewidth]{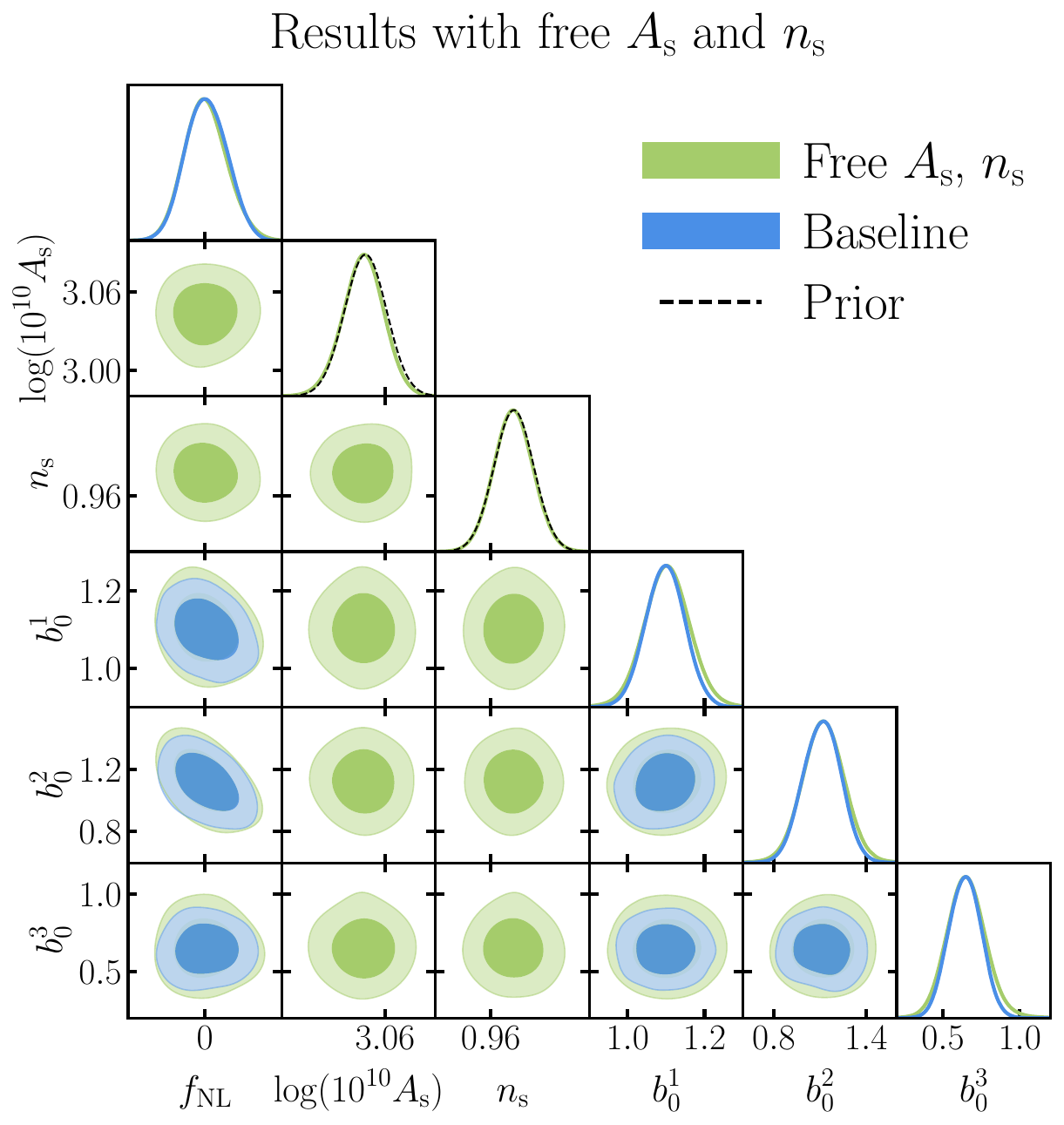}
    \caption{Contours for the default analysis, but performed with two extra free parameters: $A_\mathrm{s}$ and $n_\mathrm{s}$.}
    \label{fig:As_chain}
\end{figure}
In both cases, we still allow for a free bias amplitude in each tomographic bin. Table~\ref{tab:diff_bias} reports the marginalized posteriors for the $4$ free parameters of the analysis. 
The \fnl{} constraints are completely consistent with the baseline choice for the bias, showing deviations in the best fist values $<0.1\sigma$ and the same constraining power, showing how \fnl{} constraints are mostly insensitive to the choice of bias evolution. As expected for how the alternative bias models are set up, the bias amplitudes $b_0^i$ in those cases are compatible with $1$. 

\subsection{The impact of the primordial power spectrum}
The parameter $f_{\mathrm{NL}}$ primarily affects the largest cosmic scales, where the observed power spectrum is most sensitive to primordial fluctuations. Consequently, it can be partially degenerate with the fundamental parameters of the primordial power spectrum, namely the amplitude $A_\mathrm{s}$ and the spectral index $n_\mathrm{s}$. Allowing these parameters to vary freely improves the propagation of uncertainties and provides a more robust characterization of the constraints on $f_{\mathrm{NL}}$. In this extended analysis, we introduced $A_\mathrm{s}$ and $n_\mathrm{s}$ as additional free parameters within our differentiable likelihood framework implemented in \texttt{Blast}. This flexibility made the joint sampling of all parameters computationally efficient and straightforward. While flat priors were assumed for the standard cosmological and nuisance parameters, we applied Gaussian priors on $A_\mathrm{s}$ and $n_\mathrm{s}$ following the \Planck 2018 results~\citep{PlanckLegacy18}:
\[
A_\mathrm{s} = (2.101 \pm 0.034) \times 10^{-9}, \qquad n_\mathrm{s} = 0.9649 \pm 0.0042.
\]
The analysis employed the default configuration: \texttt{regressis} linear weights with $p = 1.6$, one bias amplitude per tomographic bin, and no application of the optimal weighting scheme. The resulting posterior distributions are shown in Fig.~\ref{fig:As_chain}, and the corresponding parameter constraints are reported in Table~\ref{tab:As_tab}.
\begin{table}[h!]
    \centering
    \renewcommand{\arraystretch}{1.3} % increases row height
    \setlength{\tabcolsep}{12pt} % increases column spacing
    \begin{tabular}{lc}
    \toprule
    Parameter & 68\% limits \\
    \midrule
    $f_{\mathrm{NL}}$ & $1^{+33}_{-36}$ \\
    $A_\mathrm{s}$    & $\left(2.098 \pm 0.032\right) \cdot 10^{-9}$ \\
    $n_\mathrm{s}$    & $0.9648 \pm 0.0040$ \\
    $b_0^1$   & $1.101 \pm 0.074$ \\
    $b_0^2$    & $1.12 \pm 0.13$ \\
    $b_0^3$    & $0.66 \pm 0.14$ \\
    \bottomrule
    \end{tabular}
    \caption{Constraints on the parameters obtained in our fiducial analysis, but with two extra free parameters: $A_\mathrm{s}$ and $n_\mathrm{s}$.}
    \label{tab:As_tab}
\end{table}
As expected, enlarging the parameter space to include $A_\mathrm{s}$ and $n_\mathrm{s}$ slightly degrades the precision on $f_{\mathrm{NL}}$, increasing its uncertainty by approximately $10\%$. The best-fit value, however, remains consistent with that obtained in the fiducial analysis (Table~\ref{tab:main_results}). The same trend is observed for the bias parameters: their uncertainties broaden, but the central values agree well with those reported in Table~\ref{tab:biases_3b}. This confirms that the inclusion of primordial parameters does not introduce significant tension, and that the treatment of degeneracies between $f_{\mathrm{NL}}$ and the primordial power spectrum is under good control.

\section{Conclusions}\label{sec:conc}
In this work, we have established constraints on primordial non-Gaussianity by cross-correlating the \Planck PR4 CMB lensing maps with quasars from DESI DR1 that have been spectroscopically confirmed. Using scales $4<\ell<300$, we measure the parameter \fnl{}, which encodes information about local-type primordial non-Gaussianity. We focus on the cross-correlation signal, as it is more robust to systematic errors, whose presence only increases the noise of the measurement. While the quasar autocorrelation does not exhibit significant signs of contamination (Fig.~\ref{fig:clgg_bin1_corrected}), such as the large-scale excess power previously observed in the DESI Legacy Survey photometric quasar sample \citep{krolewski2024constraining}, for this work we restrict the analysis to $C_\ell^{\kappa g}$, with plans to incorporate $C_\ell^{gg}$ in future data releases. To ensure that the systematic weights applied to the quasars do not remove true clustering signal, we tested them extensively on noiseless mocks (Sec.~\ref{sec:sys_weights}).

In our fiducial setup, we use the \texttt{regressis} linear weights (Sec.~\ref{sec:sys_weights}), adopt $p=1.6$, and fit simultaneously for $f_{\mathrm{NL}}$ and a free bias amplitude $b_0^i$ for each of the three tomographic redshift bins. This analysis yields \( f_{\mathrm{NL}} = 2^{+28}_{-34},\) while for $p=1.0$, we find \(f_{\mathrm{NL}} = 6^{+20}_{-24}.\) In comparison to the analysis of photometric quasars from the DESI Legacy Survey \citep{krolewski2024constraining, Dey19}, our constraints are $\sim 35\%$ tighter. The improvement can be attributed to three key factors: the use of a spectroscopic quasar sample, which is purer and less affected by systematics; the employment of \Planck PR4 lensing maps that exhibit $\approx 20\%$ lower noise compared to the 2018 release; and the application of a new catalog-based estimator (refer to Sec.~\ref{sec:ps}), which effectively avoids numerical instabilities and pixelization complications \citep{lizancos2024harmonic, wolz2025catalog}. Furthermore, while the quasar auto-correlation $C_\ell^{gg}$ in \citep{krolewski2024constraining} was significantly contaminated by systematics, we detect no excess large-scale power in the DESI DR1 sample. As a result, we do not need any additional noise modeling in the covariance matrix. 

A similar analysis to the one presented here was previously performed using the \emph{Quaia} quasar sample \citep{fabbian2025} and combining $C_\ell^{\kappa g}$ and $C_\ell^{gg}$ to obtain $f_{\mathrm{NL}}=-28^{+26}_{-24}$ for $p=1.6$ and $f_{\mathrm{NL}}=-20^{+19}_{-18}$ for $p=1.0$, and $f_\mathrm{NL}=-13.8^{+26.7}_{-25}$ when using the cross-correlation alone (and $p=1$). \emph{Quaia} is a photometric catalog of $1.3$ million quasars covering nearly the full sky. Notably, our DESI DR1 spectroscopic sample, which only covers $\sim 17\%$ of the sky and relies exclusively on $C_\ell^{\kappa g}$, achieves comparable constraining power. This highlights the impressive constraining power of DESI, which will only grow with upcoming data releases.
At the same time, our constraints are weaker than the tightest bounds from LSS, which come from the $3$D power spectrum of DESI DR1 QSOs and LRGs, which yield $f_{\mathrm{NL}}=-3\pm 9$ \citep{chaussidon2024}.
\begin{figure}[h!]
    \includegraphics[width=\linewidth]{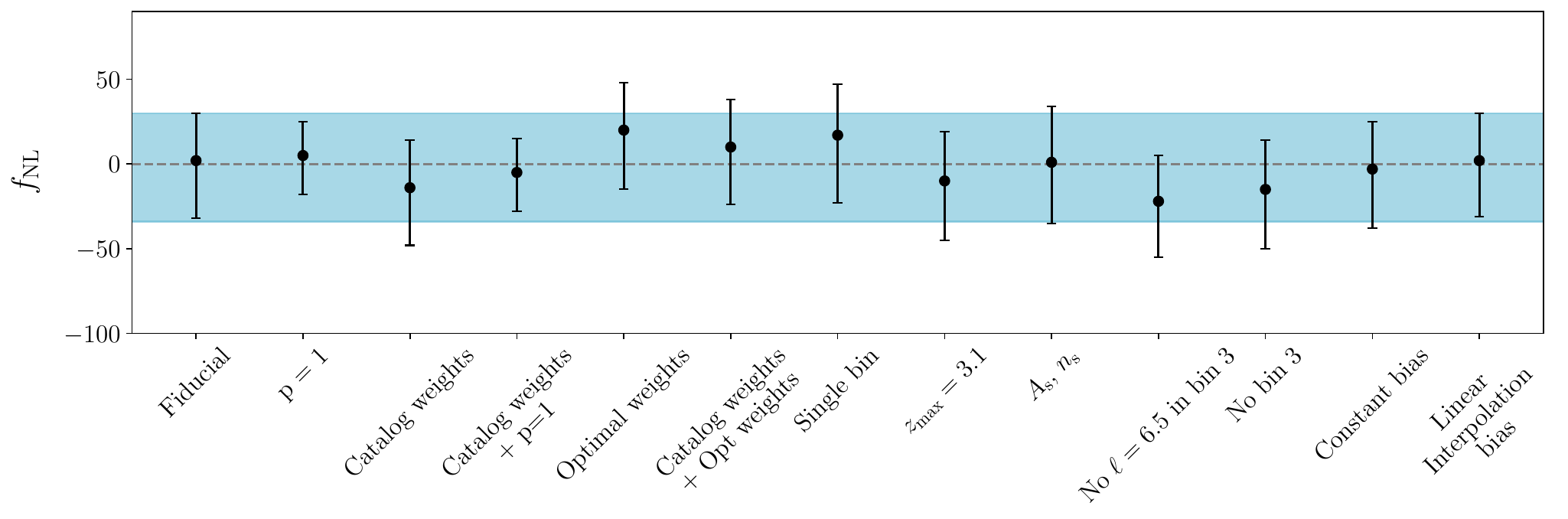}
    \caption{Constraints on \fnl{} obtained in our fiducial analysis (shaded horizontal band), and adopting alternative analysis strategies that test the robustness of our results. The numerical values shown here are listed in Table~\ref{tab:main_results}-\ref{tab:1biasfit} and described in Sec.~\ref{sec:results}.}
    \label{fig:comparison}
\end{figure}

In Sec.~\ref{sec:optw}, we derived the optimal weights for $C_\ell^{\kappa g}$ to maximize the \fnl{} signal. This optimal weighting scheme improves the constraints by $5-10\%$, similar but somewhat smaller than the gain typically achieved in three-dimensional power spectrum analyses. Furthermore, the best fits shift by $\sim 0.5-1\sigma$. Tests with a synthetic data vector and alternative tomographic binning (Sec.~\ref{sec:optw_tests}) suggest that most of the additional information from the redshift evolution of the PNG signal is already captured by splitting the sample into tomographic bins, which is the baseline configuration of this analysis. Finally, we test the impact of freeing the primordial power spectrum parameters $A_\mathrm{s}$ and $n_\mathrm{s}$, finding that our \fnl{} constraints remain completely consistent.

We presented a cosmological analysis of the large-scale clustering of DESI DR1 spectroscopically confirmed quasars, aimed at constraining the \fnl{} parameter, while a full cosmological interpretation of the sample is provided in the companion paper \citep{de2025cosmology}. Our constraints on PNG, which proved to be very stable against many analysis variations (see Fig.~\ref{fig:comparison}), are statistically weaker than those from three-dimensional power-spectrum analyses, as the measurement remains noise-dominated on the large scales that carry most of the information. However, this situation will improve significantly with future DESI data releases, which will offer greater statistical power, cover a larger sky area, and access additional scales where the \fnl{} signal is strongest. The inclusion of the quasar auto-correlation will further enhance constraining power, provided that imaging systematics are well understood and under control.

\section*{Data Availability}
The data used in this work are publicly available as part of DESI Data Release 1 (see \url{https://data.desi.lbl.gov/doc/releases/}). The data points corresponding to the figures, as well as the chains required to reproduce the corner plots, are available on Zenodo at \url{https://zenodo.org/records/19471607}. All codes and packages employed in this analysis are publicly accessible and are referenced throughout the paper.

\section*{Acknowledgments}
The authors are grateful to Emanuele Castorina for his insightful feedback and for providing the derivation that helped refine the optimal weighting scheme presented in this work.
WP acknowledges support from the Natural Sciences and Engineering Research Council of Canada (NSERC), [funding reference number RGPIN-2025-03931] and from the Canadian Space Agency.
Research at Perimeter Institute is supported in part by the Government of Canada through the Department of Innovation, Science and Economic Development Canada and by the Province of Ontario through the Ministry of Colleges and Universities.
This research was enabled in part by support provided by Compute Ontario (computeontario.ca) and the Digital Research Alliance of Canada (alliancecan.ca).

This material is based upon work supported by the U.S. Department of Energy (DOE), Office of Science, Office of High-Energy Physics, under Contract No. DE–AC02–05CH11231, and by the National Energy Research Scientific Computing Center, a DOE Office of Science User Facility under the same contract. Additional support for DESI was provided by the U.S. National Science Foundation (NSF), Division of Astronomical Sciences under Contract No. AST-0950945 to the NSF's National Optical-Infrared Astronomy Research Laboratory; the Science and Technology Facilities Council of the United Kingdom; the Gordon and Betty Moore Foundation; the Heising-Simons Foundation; the French Alternative Energies and Atomic Energy Commission (CEA); the National Council of Humanities, Science and Technology of Mexico (CONAHCYT); the Ministry of Science, Innovation and Universities of Spain (MICIU/AEI/10.13039/501100011033), and by the DESI Member Institutions: \url{https://www.desi.lbl.gov/collaborating-institutions}.

The DESI Legacy Imaging Surveys consist of three individual and complementary projects: the Dark Energy Camera Legacy Survey (DECaLS), the Beijing-Arizona Sky Survey (BASS), and the Mayall z-band Legacy Survey (MzLS). DECaLS, BASS and MzLS together include data obtained, respectively, at the Blanco telescope, Cerro Tololo Inter-American Observatory, NSF’s NOIRLab; the Bok telescope, Steward Observatory, University of Arizona; and the Mayall telescope, Kitt Peak National Observatory, NOIRLab. NOIRLab is operated by the Association of Universities for Research in Astronomy (AURA) under a cooperative agreement with the National Science Foundation. Pipeline processing and analyses of the data were supported by NOIRLab and the Lawrence Berkeley National Laboratory. Legacy Surveys also uses data products from the Near-Earth Object Wide-field Infrared Survey Explorer (NEOWISE), a project of the Jet Propulsion Laboratory/California Institute of Technology, funded by the National Aeronautics and Space Administration. Legacy Surveys was supported by: the Director, Office of Science, Office of High Energy Physics of the U.S. Department of Energy; the National Energy Research Scientific Computing Center, a DOE Office of Science User Facility; the U.S. National Science Foundation, Division of Astronomical Sciences; the National Astronomical Observatories of China, the Chinese Academy of Sciences and the Chinese National Natural Science Foundation. LBNL is managed by the Regents of the University of California under contract to the U.S. Department of Energy. The complete acknowledgments can be found at \url{https://www.legacysurvey.org/}.
Any opinions, findings, and conclusions or recommendations expressed in this material are those of the author(s) and do not necessarily reflect the views of the U. S. National Science Foundation, the U. S. Department of Energy, or any of the listed funding agencies.
The authors are honored to be permitted to conduct scientific research on I'oligam Du'ag (Kitt Peak), a mountain with particular significance to the Tohono O’odham Nation.

\clearpage
\bibliographystyle{JHEP}
\bibliography{main}
\clearpage
\appendix
\section{Derivation of the optimal $f_{\mathrm{NL}}$ weights for $C_\ell^{\kappa g}$}
\label{app:opt_weights_derivation}

In this appendix, we derive the optimal weighting scheme for the cross-correlation between a galaxy field and a matter-like tracer, such as the CMB lensing convergence $\kappa$, by generalizing the original derivation in \citep{Castorina}. 

We begin by considering the real-space galaxy overdensity field, $\delta_{g}(\mathbf{x}_1)$, and the real-space matter overdensity field, $\delta_{m}(\mathbf{x}_2)$. Assuming that the covariance matrices are diagonal in pixel space, we can write their inverses as $C_g^{-1} = 1/\sigma_g^2$ and $C_m^{-1} = 1/\sigma_m^2$ for the galaxy and matter fields, respectively (see the discussion around Eq.~2.8 in \citep{Castorina}).
Then, the optimal quadratic estimator $q$ for $f_{\mathrm{NL}}$ is:
\begin{equation}
    q = \frac{1}{2} \frac{\delta_{g}(\mathbf{x}_1)}{\sigma_g^2} \left( \frac{\partial}{\partial f_{\mathrm{NL}}} \langle \delta_{g}(\mathbf{x}_1) \delta_{m}(\mathbf{x}_2) \rangle \Big|_{f_{\mathrm{NL}}=0} \right) \frac{\delta_{m}(\mathbf{x}_2)}{\sigma_m^2}.
\end{equation}
The derivative of the cross-correlation signal with respect to $f_{\mathrm{NL}}$ is given by:
\begin{equation}
    \left( \frac{\partial \langle \delta_{g}(\mathbf{x}_1) \delta_{m}(\mathbf{x}_2) \rangle}{\partial f_{\mathrm{NL}}} \Big|_{f_{\mathrm{NL}}=0} \right) = \int \frac{d^{3}k}{(2\pi)^{3}} \frac{D(z_{1})D(z_{2})}{D(z_{1})} (b(z_{1})-p) \tilde{\alpha}(k) P_{m}(k,z_{0}) e^{i\mathbf{k} \cdot (\mathbf{x}_2 - \mathbf{x}_1)},
\end{equation}
where $P_m(k, z_0)$ is the matter power spectrum at a reference redshift $z_0$, and the transfer function $\tilde{\alpha}(k)$ is defined as:
\begin{equation}
    \tilde{\alpha}(k) \equiv \frac{3\Omega_{m} H_{0}^{2} \delta_{c}}{c^{2} k^{2} T(k)}.
\end{equation}
We can sum over all positions in the sky to obtain:
\begin{equation}
    q=\frac{1}{2} \int \frac{d^3 k}{(2 \pi)^3} \tilde{\alpha}(k) P_m\left(k, z_0\right) \int d^3 x_1 e^{-i \mathbf{k} \cdot \mathbf{x}_1} \frac{\delta_g\left(\mathbf{x}_1\right)}{\sigma_g^2}\left(b\left(z_1\right)-p\right) \int d^3 x_2 e^{i \mathbf{k} \cdot \mathbf{x}_2} \frac{\delta_m\left(\mathbf{x}_2\right)}{\sigma_m^2} D\left(z_2\right) .
\end{equation}
At this stage, we could insert the kernels $W_g$ and $W_\kappa$. We tested the impact of adding those kernels to the weights and they did not impact the constraining power, so we decided to not include them in the final weighting scheme.
Finally, one can apply the plane-wave expansion, expand the Legendre polynomials in spherical harmonics and integrate over $\mathrm{d}\Omega_k/4\pi$ to obtain:
\begin{align}
    q = \frac{4\pi}{2} \int \frac{dk k^{2}}{2\pi^{2}} \tilde{\alpha}(k) P_{m}(k, z_{0}) & \sum_{\ell,m} \left[ \int dx_{1} x_{1}^{2} j_{\ell}(kx_{1}) (b(z_{1})-p) \int d\Omega_{1} \frac{\delta_{g}(\mathbf{x}_1)}{\sigma_g^2} Y_{\ell m}(\hat{\mathbf{x}}_1) \right] \nonumber \\
    & \times \left[ \int dx_{2} x_{2}^{2} j_{\ell}(kx_{2}) D(z_{2}) \int d\Omega_{2} \frac{\delta_{m}(\mathbf{x}_2)}{\sigma_m^2} Y_{\ell m}^*(\hat{\mathbf{x}}_2) \right].
\end{align}
This expression demonstrates that the optimal measurement of $f_{\mathrm{NL}}$ is obtained by applying a redshift-dependent weight of $w_g(z) \propto (b(z)-p)$ to the galaxy field and a weight of $w_m(z) \propto D(z)$ to the matter field, results that are consistent with \citep{Castorina, Mueller19}. % in the limits $f\rightarrow0$, $b\rightarrow1$.

In our specific analysis, the matter field is provided by the CMB lensing convergence $\kappa$, which is a projected field integrated over a broad range of redshifts. Because we lack 3D information for the CMB lensing field $\kappa$, we cannot apply the optimal $D(z)$ weight to it. In conclusion, the resulting weights used in this work for $C_\ell^{\kappa g}$ are:
\begin{equation}
    w_{\mathrm{opt}}(z) = (b(z)-p) w_{\mathrm{FKP}},
\end{equation}
where $w_{\mathrm{FKP}}$ is the standard inverse-noise weighting, which is obtained from the inverse galaxy covariance $C_g^{-1}$ \citep{Castorina}.

\section{Restricting to quasars with $z<3.1$}\label{app:2}
The best \fnl{} measurement from LSS available to date was performed in \citep{chaussidon2024}. In that analysis, only the DESI quasars in the range $0.8<z<3.1$ were considered. This excludes $\approx 32.500$ quasars from the sample. That choice was motivated by the fact that the covariance matrix is estimated from simulations, the \texttt{EZMocks}, which have a simulation box that is sufficient to emulate the quasars up to $z_\mathrm{max}=3.1$ without repeating the box. 
\begin{figure}[h!]
    \includegraphics[width=\linewidth]{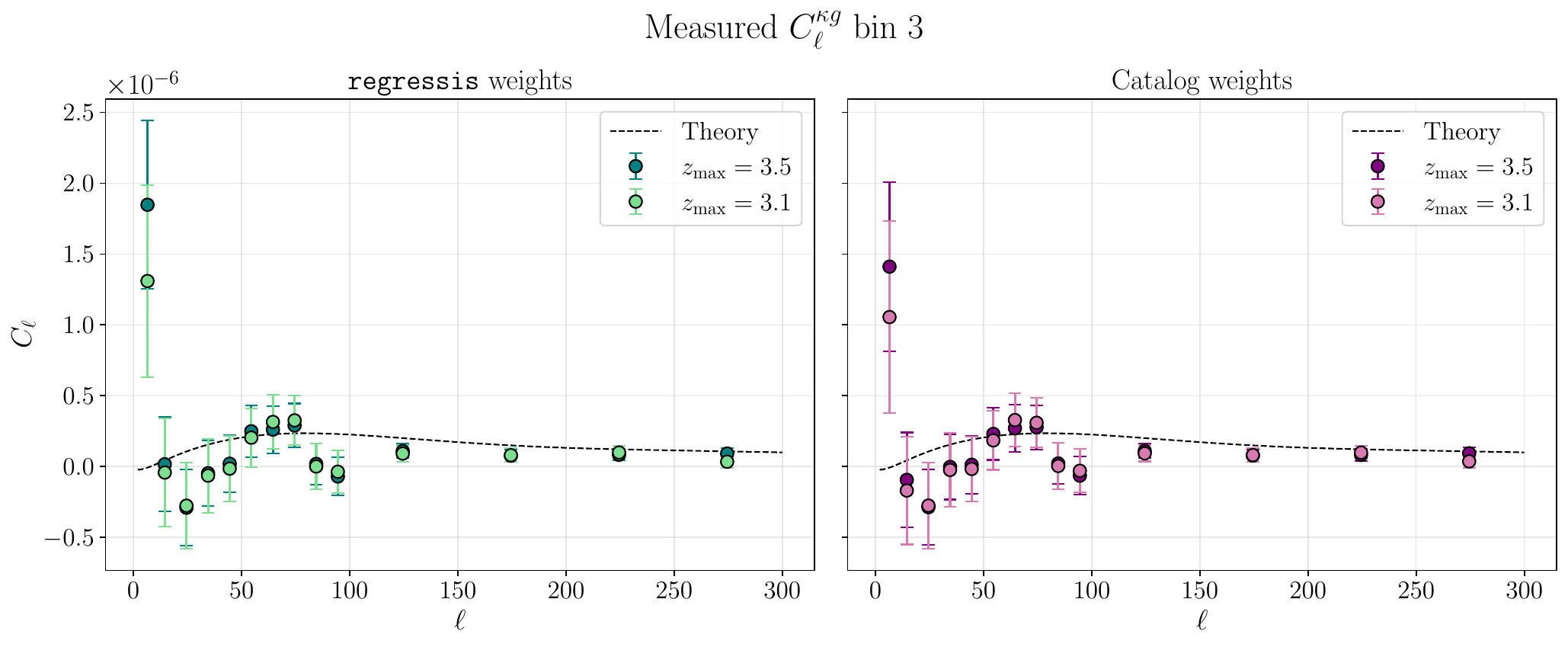}
    \caption{Measured angular cross-power spectrum $C_\ell^{\kappa g}$ in the third tomographic bin, changing the maximum redshift. \emph{Left}: measurement performed with the default weighting scheme, the \texttt{regressis} linear weights. \emph{Right}: Same measurement, but the systematic weights are the linear weights implemented in the catalog. See Sec.~\ref{sec:sys_weights} for a more detailed explanation of the weights.}
    \label{fig:cl_zmax_31}
\end{figure}
We test the impact of this choice to our analysis: we keep the default configuration of $3$ tomographic bins, fit a bias amplitude for each bin and we do not apply optimal weights. This analysis change affects the measurement, theory curve and therefore the covariance matrix, which is re-computed and tuned following the procedure outlined in Sec.~\ref{sec:ps}. As in the previous cases, results are presented for variation of systematic weights choice and values of the parameter $p$ in Table~\ref{tab:zmax31}.
\begin{table}[h!]
    \centering
    \renewcommand{\arraystretch}{1.3} % increases row height
    \setlength{\tabcolsep}{12pt} % increases column spacing
    \begin{tabular}{|c|c|c|}
        \hline
        $p$ & \texttt{regressis} linear weights & catalog linear weights\\
        \hline\hline
        $1.6$ & $f_{\mathrm{NL}} = -10^{+29}_{-35}\,\, (\chi^2=48)$  & $f_{\mathrm{NL}} = -23^{+28}_{-34}\,\, (\chi^2=48)$ \\
        \hline
        $1$ & $f_{\mathrm{NL}} = -1^{+20}_{-24}\,\, (\chi^2=48)$  & $f_{\mathrm{NL}} = -9^{+20}_{-23}\,\, (\chi^2=48)$ \\
        \hline
    \end{tabular}
    \caption{Results for the analysis performed with a value of $z_\mathrm{max}=3.1$ instead of $z_\mathrm{max}=3.5$.}
    \label{tab:zmax31}
\end{table}
Removing the higher redshift part of the sample does not affect the error bars, but has an effect on the best fit values, which move around of $\approx 1/3\,\sigma$ in the fiducial scenario. It is interesting to look at the effect on the bias amplitudes: the first two bins do not show a relevant shift, with the best fit values being completely compatible with the ones showed in Table~\ref{tab:biases_3b}. However, the best fit amplitude in the third bin is $b_3 = 0.62\pm0.13$, a value shift of $\approx 0.5\,\sigma$. Finally, the $\chi^2$ in this case is lower than in the default scenario ($48$ vs $56$). 
The difference is driven by the first multipole ($\ell=6.5$) in the third tomographic bin, as shown in Fig.~\ref{fig:cl_zmax_31}. Excluding the higher redshift part of the quasar sample causes the measurement to shift by $\approx 1\,\sigma$, with a significant impact of the $\chi^2$ value. The reasons for this shift are unknown and currently under investigation. From the figure, it is also possible to see why a lower bias is preferred, in fact, the measurement in the last multipole ($\ell=275$) is lower for both weighting schemes, explaining the preference for a lower bias value. 
\section{Affiliations}
\label{app:affiliations}
\textsuperscript{4}Lawrence Berkeley National Laboratory, 1 Cyclotron Road, Berkeley, CA 94720, USA \\
\textsuperscript{5}Department of Physics, University of California, Berkeley, 366 LeConte Hall MC 7300, Berkeley, CA 94720-7300, USA\\
\textsuperscript{6}Department of Physics, Boston University, 590 Commonwealth Avenue, Boston, MA 02215 USA\\
\textsuperscript{7}Dipartimento di Fisica ``Aldo Pontremoli'', Universit\`a degli Studi di Milano, Via Celoria 16, I-20133 Milano, Italy\\
\textsuperscript{8}INAF-Osservatorio Astronomico di Brera, Via Brera 28, 20122 Milano, Italy\\
\textsuperscript{9}Department of Physics \& Astronomy, University College London, Gower Street, London, WC1E 6BT, UK\\
\textsuperscript{10}Department of Physics and Astronomy, The University of Utah, 115 South 1400 East, Salt Lake City, UT 84112, USA\\
\textsuperscript{11}Instituto de F\'{\i}sica, Universidad Nacional Aut\'{o}noma de M\'{e}xico,  Circuito de la Investigaci\'{o}n Cient\'{\i}fica, Ciudad Universitaria, Cd. de M\'{e}xico  C.~P.~04510,  M\'{e}xico\\
\textsuperscript{12}University of California, Berkeley, 110 Sproul Hall \#5800 Berkeley, CA 94720, USA\\
\textsuperscript{13}Institut de F\'{i}sica d’Altes Energies (IFAE), The Barcelona Institute of Science and Technology, Edifici Cn, Campus UAB, 08193, Bellaterra (Barcelona), Spain\\
\textsuperscript{14}Departamento de F\'isica, Universidad de los Andes, Cra. 1 No. 18A-10, Edificio Ip, CP 111711, Bogot\'a, Colombia\\
\textsuperscript{15}Observatorio Astron\'omico, Universidad de los Andes, Cra. 1 No. 18A-10, Edificio H, CP 111711 Bogot\'a, Colombia\\
\textsuperscript{16}Institut d'Estudis Espacials de Catalunya (IEEC), c/ Esteve Terradas 1, Edifici RDIT, Campus PMT-UPC, 08860 Castelldefels, Spain\\
\textsuperscript{17}Institute of Cosmology and Gravitation, University of Portsmouth, Dennis Sciama Building, Portsmouth, PO1 3FX, UK\\
\textsuperscript{18}Institute of Space Sciences, ICE-CSIC, Campus UAB, Carrer de Can Magrans s/n, 08913 Bellaterra, Barcelona, Spain\\
\textsuperscript{19}University of Virginia, Department of Astronomy, Charlottesville, VA 22904, USA\\
\textsuperscript{20}Fermi National Accelerator Laboratory, PO Box 500, Batavia, IL 60510, USA\\
\textsuperscript{21}Institut d'Astrophysique de Paris. 98 bis boulevard Arago. 75014 Paris, France\\
\textsuperscript{22}IRFU, CEA, Universit\'{e} Paris-Saclay, F-91191 Gif-sur-Yvette, France\\
\textsuperscript{23}Center for Cosmology and AstroParticle Physics, The Ohio State University, 191 West Woodruff Avenue, Columbus, OH 43210, USA\\
\textsuperscript{24}Department of Physics, The Ohio State University, 191 West Woodruff Avenue, Columbus, OH 43210, USA\\
\textsuperscript{25}The Ohio State University, Columbus, 43210 OH, USA\\
\textsuperscript{26}Department of Physics, University of Michigan, 450 Church Street, Ann Arbor, MI 48109, USA\\
\textsuperscript{27}University of Michigan, 500 S. State Street, Ann Arbor, MI 48109, USA\\
\textsuperscript{28}Department of Physics, The University of Texas at Dallas, 800 W. Campbell Rd., Richardson, TX 75080, USA\\
\textsuperscript{29}NSF NOIRLab, 950 N. Cherry Ave., Tucson, AZ 85719, USA\\
\textsuperscript{30}Department of Physics and Astronomy, University of California, Irvine, 92697, USA\\
\textsuperscript{31}Sorbonne Universit\'{e}, CNRS/IN2P3, Laboratoire de Physique Nucl\'{e}aire et de Hautes Energies (LPNHE), FR-75005 Paris, France\\
\textsuperscript{32}Departament de F\'{i}sica, Serra H\'{u}nter, Universitat Aut\`{o}noma de Barcelona, 08193 Bellaterra (Barcelona), Spain\\
\textsuperscript{33}Department of Astronomy, The Ohio State University, 4055 McPherson Laboratory, 140 W 18th Avenue, Columbus, OH 43210, USA\\
\textsuperscript{34}Instituci\'{o} Catalana de Recerca i Estudis Avan\c{c}ats, Passeig de Llu\'{\i}s Companys, 23, 08010 Barcelona, Spain\\
\textsuperscript{35}Department of Physics \& Astronomy and Pittsburgh Particle Physics, Astrophysics, and Cosmology Center (PITT PACC), University of Pittsburgh, 3941 O'Hara Street, Pittsburgh, PA 15260, USA\\
\textsuperscript{36}Departamento de F\'{\i}sica, DCI-Campus Le\'{o}n, Universidad de Guanajuato, Loma del Bosque 103, Le\'{o}n, Guanajuato C.~P.~37150, M\'{e}xico\\
\textsuperscript{37}Instituto Avanzado de Cosmolog\'{\i}a A.~C., San Marcos 11 - Atenas 202. Magdalena Contreras. Ciudad de M\'{e}xico C.~P.~10720, M\'{e}xico\\
\textsuperscript{38}Space Sciences Laboratory, University of California, Berkeley, 7 Gauss Way, Berkeley, CA  94720, USA\\
\textsuperscript{39}Instituto de Astrof\'{i}sica de Andaluc\'{i}a (CSIC), Glorieta de la Astronom\'{i}a, s/n, E-18008 Granada, Spain\\
\textsuperscript{40}Departament de F\'isica, EEBE, Universitat Polit\`ecnica de Catalunya, c/Eduard Maristany 10, 08930 Barcelona, Spain\\
\textsuperscript{41}Department of Physics and Astronomy, Sejong University, 209 Neungdong-ro, Gwangjin-gu, Seoul 05006, Republic of Korea\\
\textsuperscript{42}CIEMAT, Avenida Complutense 40, E-28040 Madrid, Spain\\
\textsuperscript{43}Department of Physics \& Astronomy, Ohio University, 139 University Terrace, Athens, OH 45701, USA\\
\textsuperscript{44}National Astronomical Observatories, Chinese Academy of Sciences, A20 Datun Road, Chaoyang District, Beijing, 100101, P.~R.~China
\end{document}